\documentclass[aps,pra,twocolumn,superscriptaddress]{revtex4-1}
\usepackage{epsfig}
\usepackage{amsmath,amssymb,amsfonts}
\usepackage{graphicx}
\newcommand{\beq}{\begin{equation}}
\newcommand{\eeq}{\end{equation}}
\newcommand{\nn}{\nonumber}
\renewcommand\Re{\operatorname{Re}}

\usepackage{braket}

\begin{document}
\title{Micromotion in trapped atom-ion systems}
\author{\surname{L\^e} Huy Nguy\^en}
\affiliation{NUS Graduate School for Integrative Science and Engineering, 28 Medical Drive, 117456, Singapore}
\affiliation{Centre for Quantum Technologies, National University of Singapore, 3 Science Drive 2, 117543, Singapore}
\author{Amir \surname{Kalev}}
\affiliation{Centre for Quantum Technologies, National University of Singapore, 3 Science Drive 2, 117543, Singapore}
\author{Murray D. \surname{Barrett}}
\affiliation{Centre for Quantum Technologies, National University of Singapore, 3 Science Drive 2, 117543, Singapore}
\author{Berthold-Georg \surname{Englert}}
\affiliation{Centre for Quantum Technologies, National University of Singapore, 3 Science Drive 2, 117543, Singapore}
\affiliation{Department of Physics, National University of Singapore, 2 Science Drive 3, 117542, Singapore}

\begin{abstract}

We examine the validity of the harmonic approximation, where the radio-frequency ion trap is treated as a harmonic trap, in the problem regarding the controlled collision of a trapped atom and a single trapped ion. This is equivalent to studying the effect of the micromotion since this motion must be neglected for the trapped ion to be considered as a harmonic oscillator. By applying the transformation of Cook and Shankland we find that the micromotion can be represented by two periodically oscillating operators. In order to investigate the effect of the micromotion on the dynamics of a trapped atom-ion system, we calculate (i) the coupling strengths of the micromotion operators by numerical integration and (ii) the quasienergies of the system by applying the Floquet formalism, a useful framework for studying periodic systems. It turns out that the micromotion is not negligible when the distance between the atom and the ion traps is shorter than a characteristic distance. Within this range the energy diagram of the system changes remarkably when the micromotion is taken into account, which leads to undesirable consequences for applications that are based on an adiabatic process of the trapped atom-ion system. We suggest a simple scheme for bypassing the micromotion effect in order to successfully implement a quantum controlled phase gate proposed previously and create an atom-ion macromolecule. The methods presented here are not restricted to trapped atom-ion systems and can be readily applied to studying the micromotion effect in any system involving a single trapped ion.
\end{abstract}
\begin{widetext}
\maketitle
\end{widetext}
\section{Introduction}

The dynamics of a trapped atom-ion system has recently been studied in great detail \cite{Idziaszek07,Idziaszek09}, and it shows an interesting application in realizing a two-qubit gate for quantum computing \cite{Doerk10}. The system is composed of a single atom trapped in a harmonic trap interacting with an ion trapped in a radio-frequency (rf) trap.

In these studies, the rapid motion of the ion on a short time-scale --- the \emph{micromotion} --- is averaged out. This procedure is referred to in the literature as the harmonic approximation since it produces an effective harmonic motion of the ion, as if the ion were trapped in a time-independent harmonic trap. In fact, many of the proposals for applications of this system are derived based on the harmonic approximation. However, there is a concern about the validity of this approximation since it is known that the kinetic energy of the micromotion is comparable to that of the ion's harmonic motion \cite{Berkeland98}. Motivated by such concerns, we present here two approaches, as mentioned in the Abstract, to studying the effect of the micromotion in trapped atom-ion systems. 

For this purpose we make use of the Floquet formalism \cite{Chu85}. The potential of the rf electric field used to trap ions depends periodically on time.   When  the  potential  is  periodic,  Floquet  theory provides a powerful tool for treating the exact dynamics of the quantum system. It enables one to compute the so-called quasienergies and quasienergy states, also referred to as Floquet states, which are the analogs of the eigenenergies and eigenstates of a time-independent system.  Hence, it offers a \emph{quantum-mechanical treatment} of the micromotion problem in any system involving a single trapped ion. By studying the exact quasienergies and Floquet states we obtain valuable information about the role of the micromotion in such systems. Its effect will be then deduced by comparing the exact dynamics, as derived by the Floquet formalism, with the approximate one given by the harmonic approximation. In particular, this comparison reflects on the validity of various proposals for applications using controlled collisions of trapped atoms and ions. Although we demonstrate the method for the system of interacting trapped atoms and ions, it can be readily applied to any system in which the trapped ion is coupled to an external time-independent subsystem.

The article is organized as follows. In Secs.~\ref{ionmicromotion} and \ref{atomionsystem} we describe the ion's micromotion and the trapped atom-ion systems. Section~\ref{floquet} is on the Floquet formalism and how we use it to study the effect of a periodically oscillating potential on the energy structure of a quantum system. In Sec.~\ref{micromotioneffect} we study the micromotion effect in a trapped atom-ion system by considering the micromotion-induced coupling and computing the exact quasienergies of the system. We also suggest a simple scheme to bypass the micromotion effect to realize some interesting applications based on interaction of trapped atoms and ions. Finally, we offer conclusions and present some technical material in the appendices.

\section{Ion micromotion and harmonic approximation}\label{ionmicromotion}

An excellent review of the classical and quantum dynamics of a single trapped ion is given in Ref. \cite{Leibfried03}. Here we briefly summarize the key points in the motion of trapped ions and explain the reasoning behind the harmonic approximation. 

For an ion in a rf trap, the equations of motion along the $x, y, z$ axes are decoupled. With $\omega$ the frequency of the oscillating potential and $m_i$ the ion's mass, the Hamiltonian for the motion along the $x$ axis is 
\begin{equation}\label{ham}
H(t)=\frac{P^2}{2m_i} + \frac{1}{8}m_i \omega^2 X^2[a+2q \cos(\omega t)],
\end{equation}
where $a$ and $q$ are trap parameters which depend on $m_i$, $\omega$, the ion's charge $Ze$, and the characteristics of the trapping potential. The Hamiltonians for the motion along the $y$ and $z$ directions assume similar forms with different values for the parameters $a$ and $q$. In a linear Paul trap, for instance, we have 
\begin{align}\label{paul}
&q_x=-q_y=q, \ q_z=0, \nn \\
&a_x=a_y=-\frac{1}{2}a_z=a,
\end{align}
which means the oscillating part of the trapping potential only exists in the radial directions $(x,y)$. Typically, $|q| \ll 1$ and $|a| \ll q^2$.

The ion's motion is a combination of a harmonic oscillation at a secular frequency $\omega_0$ and a micromotion that oscillates at the frequency $\omega$ of the driving potential. The secular frequency is 
\beq\label{sf}
\omega_0 = \frac{\omega}{2} \sqrt{a+\frac{q^2}{2}};
\eeq
thus, $\omega_0 \ll \omega$; that is, the secular motion is much slower than the micromotion. These conclusions are drawn from the approximate solution
\beq\label{ionclassical}
x(t) \approx x_0\cos(\omega_0 t)\left[1+\frac{q}{2} \cos(\omega t)\right]
\eeq
of the classical equation of motion with $x_0$ an arbitrary constant. Since the amplitude of the micromotion goes as the small parameter $q$, it can be seen as a jiggling motion around the overall path of the secular motion. When one averages over the short time period of the micromotion, the $\cos(\omega t)$ term in Eq.~\eqref{ionclassical} vanishes and the resulting motion of the ion is a harmonic oscillation. This is the so-called harmonic approximation.  
 
Cook, Shankland, and Wells provide a quantum-mechanical derivation of the harmonic approximation in Ref. \cite{Cook85}. They write the wave function of the ion as 
\beq\label{phase}
\Psi(x,t) = \exp\left[-\frac{i}{4\hbar}m_i q \omega x^2 \sin(\omega t)\right] \Phi(x,t)
\eeq
and insert this into the Schr\"{o}dinger equation for the Hamiltonian of Eq.~\eqref{ham}. It follows that the effective wave function $\Phi(x,t)$ obeys the modified Schr\"{o}dinger equation  
\begin{align}\label{mS}
i\hbar\frac{\partial}{\partial t} \Phi(x,t) = &\Big[-\frac{\hbar^2}{2m_i}\frac{\partial^2}{\partial x^2} +\frac{1}{2}m_i \omega_0^2 x^2  \nn \\
& - m_i (\gamma\omega_0)^2 x^2 \cos(2\omega t) \nn \\
&+2 i\hbar \gamma\omega_0 \left(x\frac{\partial}{\partial x}+\frac{1}{2}\right)\sin(\omega t)\Big]\Phi(x,t),
\end{align}
where $\omega_0$ is the secular frequency mentioned above and the factor $\gamma$ is 
\beq\label{gamma}
\gamma = \frac{1}{\sqrt{2\left(1+\frac{2a}{q^2}\right)}}.
\eeq 

Cook \emph{et al.} argue that most of the fast time dependence of $\Psi(x,t)$ is contained in the exponential factor and hence $\Phi(x,t)$ may be treated as a slowly varying function of time. Therefore, one may take in Eq.~\eqref{mS} the time average over the short time interval $2\pi/\omega$ of the micromotion; after that we are left with the well-known Schr\"{o}dinger equation for a harmonic oscillator. This is the quantum-mechanical basis for the harmonic approximation. Equation~\eqref{mS} tells us that this approximation is valid only when the time-dependent terms have little effect on the unperturbed wave function. As we see later, while this is all right for a single trapped ion, it may not hold when the ion is coupled to an external system.

The  Hamiltonian associated with the effective wave function $\Phi(x,t)$ can be read off from Eq.~\eqref{mS},
\beq
H_{\mathrm{eff}}(t)=\frac{P^2}{2m_i} + \frac{1}{2} m_i \omega_0^2 X^2 +H_{\mathrm{mm}}(t),
\eeq  
where
\beq\label{im}
H_{\mathrm{mm}}(t) = - m_i (\gamma\omega_0)^2 X^2 \cos(2 \omega t)- \gamma\omega_0 \{X,P\} \sin( \omega t)
\eeq 
accounts for the time-dependent contribution of the micromotion. Here $\{X,P\}$ is the anti commutator of the position and momentum operators. The form of $H_{\mathrm{eff}}$, where the time-independent part of the secular motion and the time-dependent one of the micromotion are separated, makes it more convenient to work with when one wishes to study the effect of the micromotion. In the next section we make use of this transformation to investigate the trapped atom-ion system. 
\section{Trapped atom-ion systems}\label{atomionsystem}

\begin{figure}[h]
\centering
\includegraphics[scale=0.8]{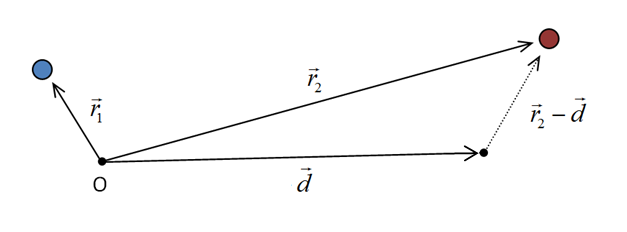}
  \caption{A trapped atom-ion system. O is the center of the atom trap which is chosen as the coordinate origin; $\vec{r}_1$ and $\vec{r}_2$ are the position vectors of the atom and ion, respectively; and $\vec{d}$ is the position vector of the center of the ion trap.}\label{figtrapatomion}
\end{figure}

The system is composed of an atom in a harmonic trap interacting with an ion in a rf trap as shown in Fig.~\ref{figtrapatomion}. To simplify the problem we first consider the one-dimensional (1D) system in which the atom and ion are confined to moving along only one axis, say the $x$ axis. The realistic three-dimensional (3D) system is discussed in Sec.~\ref{3D} below. Let us choose the origin of coordinate at the center of the atom trap. When the trap centers are separated by a distance $d$, the Hamiltonian of the system is 
\begin{align}\label{atomionH}
H(t)=&H_a+H_i (t) + V_{\mathrm{int}} \nn \\
 =&\frac{P_a^2}{2 m_a} + \frac{1}{2}m_a \omega_a^2 X_a^2+\frac{P_i^2}{2m_i}  \nn \\
&+ \frac{1}{8}m_i \omega^2 (X_i-d)^2[a+2q \cos(\omega t)]+V_{\mathrm{int}}.
\end{align}
The indices $a$ and $i$ label the atom and ion, respectively.

The properties of the interaction potential $V_{\mathrm{int}}$ are well-documented in Ref.~\cite{Idziaszek07}. When the distance $r$ between the atom and the ion is large, this potential has the form $V_{\mathrm{int}}(r)~\approx~-(\alpha e^2)/(2r^4)$, where $\alpha$ is the polarizability of the atom. This long-range form results from the attractive force between the ion's charge and the atom's induced dipole. The characteristic length of the atom-ion interaction is defined by $R^* = \sqrt{m \alpha e^2/\hbar^2}$, where $m$ is the reduced mass of the atom-ion system. This length indicates the range within which the atom-ion interaction potential is larger than the quantum kinetic energy $\hbar^2/2 m r^2$. For the $^{135}\mbox{Ba}^{+}$ and $^{87}\mbox{Rb}$ system, $R^*$ is around 5544 Bohr radii.

The behavior of the interaction potential at short distances is very different from its long-range form. The short-range potential is, in general, non-central and depends on the electronic configurations of the atom and ion. This spin dependence of the short-range potential is key to the implementation of a quantum phase gate proposed in Ref.~\cite{Doerk10} (here ``spin" means the binary alternative of two hyperfine states). Idziaszek \emph{et al.} \cite{Idziaszek07} show how to take into account the effect of the short-range potential by utilizing quantum defect theory where the short-range potential can be characterized by a single quantum defect parameter called the short-range phase . The basic idea is to replace the potential $V_{int}$ with its long-range form while imposing a specific boundary condition on the wave function at a small distance $r_{\mathrm{min}} \ll R^*$. In addition, this distance $r_{\mathrm{min}}$ must be sufficiently larger than the length scale set by the short-range potential, which is a few Bohr radii. 

To investigate the micromotion effect, we follow Cook \emph{et al.} and make the transformation 
\beq\label{trans}
\Psi(x_a,x_i,t)\! = \exp\!\left[-\frac{i}{4\hbar}m_i q \omega (x_i-d)^2\! \sin(\omega t)\right]\!\! \Phi(x_a,x_i,t).
\eeq
The resulting effective Hamiltonian for the effective wave function $\Phi(x_a,x_i,t)$ is the sum of a time-independent Hamiltonian and an oscillating term $H_{\mathrm{mm}}(t)$ that represents the ion's micromotion,
 \begin{align}\label{atomiontransH}
H(t)=&\frac{P_a^2}{2 m_a} + \frac{1}{2}m_a \omega_a^2 X_a^2 +\frac{P_i^2}{2m_i} +\frac{1}{2}m_i \omega_0^2 (X_i-d)^2 \nn \\
 &- \frac{\alpha e^2}{2(X_i - X_a)^4} + H_{\mathrm{mm}} (t), \nn \\
H_{\mathrm{mm}}(t)= &- m_i (\gamma\omega_0)^2 (X_i-d)^2 \cos(2 \omega t)  \nn \\
 &- \gamma\omega_0 \{X_i - d,P_i\} \sin( \omega t),
\end{align}
where we already replaced the interaction potential by its long-range form. Any calculation of the wave function must take into account the boundary condition dictated by the short-range phase. 

In the harmonic approximation, $H_{\mathrm{mm}}(t)$ is neglected and we only need to deal with a time-independent system. This approximate approach has been studied in great detail previously \cite{Idziaszek07,Doerk10}. Our purpose is to investigate how the micromotion affects the unperturbed system; thus, we have to work with the full Hamiltonian of Eq.~\eqref{atomiontransH}. We are particularly interested in how the state of the system evolves in an adiabatic process where the trap distance is changed slowly in time as such a process is vital for the implementation of the proposed quantum phase gate in Ref. \cite{Doerk10}. 

Although the micromotion is represented by an oscillating term similar to an electromagnetic field, it is an intrinsic property of the system and cannot be switched on or off. Therefore, time-dependent perturbation theory which emphasizes the transitions between unperturbed states is not a suitable approach to the micromotion problem, which is the reason why we need to consider the Floquet formalism.

\section{Floquet formalism}\label{floquet}
\subsection{Floquet theory}
The Floquet formalism was first considered by Shirley \cite{Shirley65} and has been developed in great depth for studying atomic and molecular multiphoton processes in intense laser fields \cite{Chu85}. Thus, we have an advanced mathematical framework ready in our hands to investigate the exact quantum dynamics of the trapped atom-ion system. In this section we outline the key points which are important for our study.

The  Hamiltonian of Eq.~\eqref{atomiontransH} satisfies the periodicity condition \mbox{$H(t+T)=H(t)$} with $T=2\pi / \omega $. According to the Floquet theorem, the Schr\"{o}dinger equation for such a periodic system,
\beq\label{periodicS}
i \hbar \frac{\partial}{\partial t} \Psi(t) = H(t) \Psi(t),
\eeq
 adopts a special class of solutions called the Floquet solutions which can be expressed in terms of the quasienergy $\epsilon$ and the Floquet wave function $u(t)$ as 
 \beq\label{Floquetsolution}
 \Psi(t)=\exp\left(-\frac{i}{\hbar} \epsilon  t\right) u(t),
 \eeq
where $u(t)$ is a function of both time and space coordinates and is periodic in time; that is,
\beq 
 u(t+T)=u(t).
 \eeq
 A substitution of $\Psi(t)$ from Eq.~\eqref{Floquetsolution} to Eq.~\eqref{periodicS} yields
\beq\label{Floqueteq}
\left[H(t) - i\hbar \frac{\partial}{\partial t}\right]u(t)=\epsilon u(t).
\eeq
Thus, the quasienergy and the Floquet wave function are respectively the eigenvector and eigenvalue of the operator
\beq
H_{\mathrm{F}}(t)=H(t) - i\hbar \frac{\partial}{\partial t},
\eeq
which is called the Floquet Hamiltonian. One observes from Eq.~\eqref{Floquetsolution}  that quasienergies and Floquet states are to a periodic system what eigenenergies and eigenstates are to a time-independent system. 

The solutions to the Floquet eigenvalue equation \eqref{Floqueteq} has the following important Brillouin-zone-like structure: If $u(t)$ is a Floquet state with quasienergy $\epsilon$, then $u(t) \exp(i k\omega t)$  is also a Floquet state with quasienergy $\epsilon +k \hbar \omega$ for any integer $k$. These sates  are physically equivalent because they belong to a unique wave function, inasmuch as
\beq
\Psi(t)= u(t)e^{-\frac{i}{\hbar} \epsilon t} = \left[u(t)e^{i k \omega t}\right] e^{-\frac{i}{\hbar} (\epsilon + k \hbar \omega)t}.
\eeq 
In other words, Floquet states and quasienergies come in congruent classes modulo $\omega$, each containing an infinite number of equivalent members. It is convenient to denote the Floquet states and their corresponding quasienergies by $u_{n,k}$ and $\epsilon_{n,k}$ where the index $n$ indicates physically different classes and $k$ different members in a class. Since for each quasienergy $\epsilon$ there are equivalent quasienergies $\epsilon +k \hbar \omega$, it is always possible to reduce an arbitrary quasienergy to a single zone $[E-\hbar \omega/2,E+ \hbar \omega/2]$ for an arbitrary energy $E$. As there can be only one member from each class $\epsilon_{n,k}$ in a single zone, we need not worry about the redundancy of the physically equivalent states if we restrict our study of the quasienergies to one zone. 

Sambe \cite{Sambe73} introduces the extended Hilbert space for all the square-integrable periodic functions with period $T$ in which the scalar product is defined as
\beq
\langle \braket{u(\vec{r},t),v(\vec{r},t)}\rangle=(1/T)\int_0^T dt \braket{u(\vec{r},t),v(\vec{r},t)}.
\eeq
This scalar product is a 4D generalization of the normal scalar product in 3D. Time and space can be treated on equal footings in the extended Hilbert space. Since the Floquet states are constant vectors in this space, an advantage of working with it is that methods developed for time-independent quantum mechanics, such as the Rayleigh-Schr\"{o}dinger perturbation method and the variational principle can be readily generalized for the Floquet states. The generalization is straightforward thanks to the similarity between the Floquet eigenvalue equation \eqref{Floqueteq} and the time-independent Schr\"{o}dinger equation.

The final properties of the Floquet states we need to mention are the orthonormality
\beq
\langle \braket{u_{n,k},u_{m,j}}\rangle=\delta_{n,m} \delta_{k,j},
\eeq
and the completeness,
\beq
\sum_{n,k} \ket{u_{n,k}}\rangle \langle \bra{u_{n,k}}=I,
\eeq
which comes from the fact that any square-integrable and $T$-periodic wave function $\psi(t)$ can be expanded as
\beq
\psi(t)=\sum_{n,k} u_{n,k}(t) \langle \braket{u_{n,k},\psi} \rangle,
\eeq
which is just a generalized Fourier series of $\psi(t)$.

Recall that without the micromotion the trapped atom-ion system has a time-independent Hamiltonian with well-documented eigenenergies and eigenstates \cite{Doerk10,Idziaszek07}. If the micromotion is included, the system is a periodic one that possesses well-defined quasienergies and Floquet states. When the strength of the micromotion $H_{\mathrm{mm}}(t)$ in Eq.~\eqref{atomiontransH} is reduced to zero, these quasienergies and Floquet states must approach the eigenenergies and eigenstates obtained by the harmonic approximation. The difference between the exact quasienergies (Floquet states) and the approximate eigenenergies (eigenstates) tells us how important the micromotion effect is. 

To be more specific, let us consider the Hamiltonian of the trapped atom-ion system which can be written as
\beq
H(t) = H_0 + H_{\mathrm{mm}}(t),
\eeq
where $H_0$ is the unperturbed, time-independent part with eigenenergies $E_{n}$ and eigenstates $\phi_{n}$. When the micromotion is taken into account these states will transform to some Floquet states $u_{n}(t)$ with quasienergies $\epsilon_{n}$. An obvious way to quantify the micromotion effect is to calculate the  energy difference 
\beq
\delta E=|\epsilon_{n}-E_{n}|
\eeq
and the deviation between the exact, oscillating probability density $\rho_{n}(t) = |u_{n}(t)|^2$ and the approximate, static density $\rho^{(0)}_{n}= |\phi_{n}|^2$. There are many ways to measure the difference between these probability densities. An example is 
\beq
\delta \rho_n = \frac{1}{T}\int dt \int dx \left|\rho_{n}(x,t)-\rho^{(0)}_{n}(x)\right|,
\eeq
which is the average over time and space of the absolute difference between $\rho_{n}(x,t)$ and $\rho^{(0)}_{n}(x)$. We see later that, for our particular problem, it suffices to calculate the energy difference; however, the oscillation of the wave function may have important effects in general.

\subsection{Floquet Hamiltonian}\label{FH}

In this section we show how to obtain the Floquet states and quasienergies by the Floquet Hamiltonian method. Let us consider a system described by the Hamiltonian
\beq\label{generalperiodicH}
H(t)=H_0 + V\cos (\omega t),
\eeq
where $V$ is a time-independent operator. The corresponding Floquet Hamiltonian for this system is
\beq
H_{\mathrm{F}}(t)=H_0 - i\hbar \frac{\partial}{\partial t} + V\cos (\omega t).
\eeq
Suppose we know the eigenenergy $E_n$ and eigenstates $\ket{n}$ of the unperturbed Hamiltonian $H_0$. Then the quasienergies and Floquet states of the \emph{unperturbed} Floquet Hamiltonian 
\beq
H_{\mathrm{F}}^{(0)}=H_0 -i\hbar \frac{\partial}{\partial t}
\eeq
are simply
\begin{align}
\epsilon_{n,k}^{(0)}&=E_n+k\hbar \omega, \nn \\
u_{n,k}^{(0)}&=\ket{n} e^{ik \omega t},
\end{align}
for any integer $k$. We may now obtain the matrix elements of the full Floquet Hamiltonian $H_{\mathrm{F}}$ by utilizing the scalar product of the extended Hilbert space,
\begin{align}\label{floquetmatrix}
\langle \bra{u^{(0)}_{n,k}}H_F\ket{u^{(0)}_{m,j}}\rangle =&\left( E_n+k\hbar \omega \right) \delta_{n,m} \delta_{k,j} \  \nn \\
&+\frac{1}{2}\langle n |V|m\rangle \left( \delta_{k,j+1} + \delta_{k,j-1} \right).
\end{align}
Diagonalizing the above infinite matrix gives the exact quasienergies and Floquet states of the periodic system. In practice one needs to truncate the Floquet Hamiltonian matrix by choosing $n=1, \ldots, N_e$ and $k=-N_f, \ldots, N_f$ for some sufficiently large numbers $N_e$ and $N_f$. Then, the linear size of the Floquet Hamiltonian matrix is \mbox{$N=N_e(2N_f+1)$}.

Besides the numerical diagonalization of the Floquet Hamiltonian, we can also use perturbation methods \cite{Sambe73} to find the approximate quasienergies and Floquet states when the oscillating potential is weak. According to the Rayleigh-Schr\"{o}dinger perturbation method generalized for Floquet states, the lowest-order corrections for the quasienergies and Floquet states are
\begin{align}
\epsilon^{(1)}_{n,k}&=\langle \bra{u^{(0)}_{n,k}}V(t)\ket{u^{(0)}_{n,k}}\rangle, \nn \\
u^{(1)}_{n,k}&=\sum_{\substack{m,j\\ \{m,j\}\neq\{n,k\}}}\frac{\langle
\bra{u^{(0)}_{m,j}}V(t)\ket{u^{(0)}_{n,k}}\rangle}{\epsilon^{(0)}_{n,k} - \epsilon^{(0)}_{m,j}}u^{(0)}_{m,j},\nn \\
\epsilon^{(2)}_{n,k}&=\sum_{\substack{m,j\\ \{m,j\}\neq\{n,k\}}}\frac{\left|\langle \bra{u^{(0)}_{m,j}}V(t)\ket{u^{(0)}_{n,k}}\rangle \right|^2}{\epsilon^{(0)}_{n,k} - \epsilon^{(0)}_{m,j}}.
\end{align}
Upon inserting $V(t)=V\cos {\omega t}$ and $\epsilon^{(0)}_{m,j}=E_m + j \hbar \omega$ into the above equations and evaluating the scalar products, we arrive at
\begin{align}
\epsilon^{(1)}_{n,k}&=0, \nn \\
u^{(1)}_{n,k}&=e^{ik\omega t}\!\sum_{m}\! \ket{m}\!\bra{m}V\ket{n}\!\frac{\Delta E_{nm}\!\cos (\omega t)\!+ i \hbar \omega \sin (\omega t)}{{\Delta E^2_{nm}} - (\hbar\omega)^2}, \nn \\
\epsilon^{(2)}_{n,k}&=\sum_{m} \frac{1}{2}\left|\bra{m}V\ket{n}\right|^2 \frac{{\Delta E_{nm}}}{{\Delta E^2_{nm}} - (\hbar\omega)^2},
\end{align}
where $\Delta E_{nm}=E_n - E_m$. 

The fact that the energy shift is the same for all values of $k$ implies that the Rayleigh-Schr\"{o}dinger perturbation method preserves the $\omega$-modulo structure of each class of quasienergies. It is clear from the above equation that this perturbation approach is valid only when 
\beq
\left|\bra{m}V\ket{n}\right|^2 \ll |(E_n-E_m)^2 - (\hbar \omega)^2| ,
\eeq 
which requires that no resonance exists between $\ket{n}$ and any other state $\ket{m}$, that is, $|E_n - E_m|\not= \hbar \omega$.
When there are two eigenstates close to resonance, one needs to use the almost-degenerate perturbation method which is discussed in the next section.

\subsection{Resonance}\label{resonances}
\begin{figure}[t]
\centering
\includegraphics[scale=0.35]{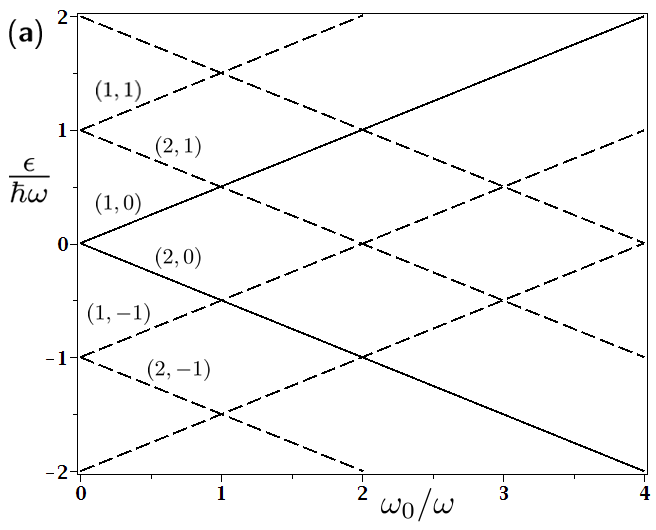}
\includegraphics[scale=0.34]{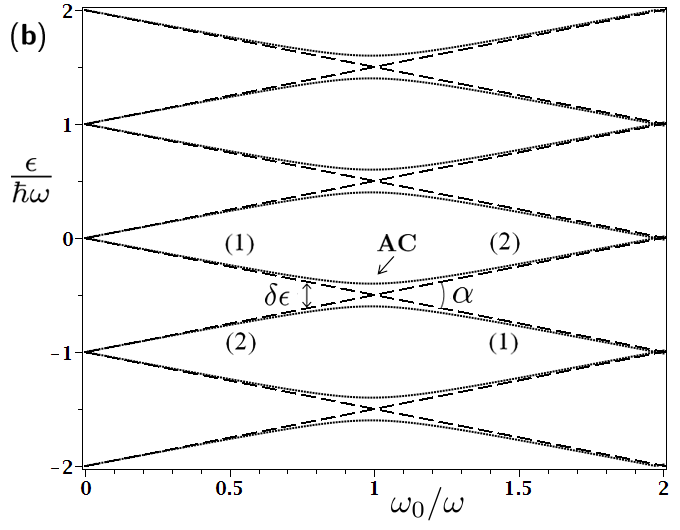}
  \caption{(a) Eigenenergies and unperturbed quasienergies. The solid lines indicate the two eigenenergy levels, the dash ones show the zone structure of the unperturbed quasienergies. (b) A comparison of the exact quasienergies, represented by the dotted curves, with the unperturbed quasienergies. A typical avoided crossing caused by the oscillating field is marked by AC.}\label{fig2level}
\end{figure}

To demonstrate clearly the important role of resonance in quasienergy structures, we first discuss a two-level system whose Hamiltonian has the form in Eq.~\eqref{generalperiodicH} with
\beq
H_0=\frac{\hbar \omega_0}{2}\begin{pmatrix}  1 &0 \\ 0 &-1 \end{pmatrix} ,
\eeq
and
\beq
V=\hbar  \eta\begin{pmatrix} 0 &1 \\ 1 &0 \end{pmatrix},
\eeq
where the coupling constant $\eta$ is small enough so that perturbation theory is applicable. Furthermore, we allow the frequency $\omega_0$ to be varied from zero to well beyond the frequency $\omega$ of the oscillating field. This system is exactly solvable but we choose perturbation methods as they are still useful when one goes to multilevel systems. The eigenenergy diagram and the corresponding quasienergy diagram of the unperturbed system is shown in Fig.~\ref{fig2level}(a), where each quasienergy level is indicated by its double index $\{n,k\}$. Note that resonances appear as apparent crossings in the unperturbed quasienergy diagram, here drawn in four zones from $-2\omega$ to $2\omega$.

Now we turn to the calculation of the quasienergies of the full system. To gain insight we first use the perturbation method to find the approximate quasienergies and Floquet states. Far from resonance, when 
\[\left|\frac{\omega_0}{\omega}-1\right| \gtrsim 1/2, \]
one may use the Rayleigh-Schr\"{o}dinger perturbation theory to obtain
\begin{align}\label{offresqe}
\epsilon_{n,k}&\approx \epsilon^{(0)}_{n,k} +\epsilon^{(2)}_{n,k} =  (-1)^{n+1}\frac{\hbar\omega_0}{2}\left(1+\frac{\eta^2}{\omega_0^2-\omega^2}\right),
\end{align}
and
\begin{align}\label{offresFs}
u_{n,k}\approx &u^{(0)}_{n,k}+u^{(1)}_{n,k} \nn \\
=& \ket{n}e^{ik\omega t}\nn \\
&+\eta \ket{\bar{n}}e^{ik\omega t}\frac{(-1)^{n+1} \omega_0\cos (\omega t)+i\omega \sin (\omega t)}{\omega_0^2 - \omega^2},
\end{align}
with $n=1,2$ and $\bar{n}=\left[3-(-1)^n\right]/2$. Thus, the energy shift is of the order
\beq
\delta \epsilon = \epsilon_{n,k} - \epsilon^{(0)}_{n,k}\sim \hbar \omega_0 \left(\frac{\eta}{\omega}\right)^2,
\eeq 
which is very small if $\eta\ll \omega$. The small periodic correction to the Floquet state means that this state is represented by a wave function that undergoes small oscillations around the mean value which is the unperturbed wave function. Therefore, the unperturbed wave function is a good approximation to the Floquet state in this far off-resonance region.

In the opposite extreme, where the system is close to resonance when $\omega_0/\omega\approx 1$, one needs to use the almost-degenerate perturbation method \cite{Sambe73}. We start with computing the Floquet Hamiltonian using Eq.~\eqref{floquetmatrix}. If one arranges the double index $\{n,k\}$ such that $n$ goes through $1,2$ before each change in $k$, the Floquet Hamiltonian has this infinite block-tridiagonal structure:
\begin{widetext}
\begin{align} \bordermatrix{ &\ & {-2} &{-1} &{0} &{1} &{j=2}  \cr
 &. &. &. &. &. &. &. \cr
                  {-2} &. &{H_0\! -\! 2\hbar\omega I} &\frac{V}{2} &0 &0 &0 &. \cr
                   {-1} &. &\frac{V}{2}  &{H_0\! -\! \hbar\omega I} &\frac{V}{2}  &0 &0 &. \cr
                    {\  \ 0} &. &0 &\frac{V}{2}  &H_0 &\frac{V}{2}  &0 &.\cr
                   {\  \ 1} &. &0 &0 &\frac{V}{2}  &H_0\!+\!\hbar\omega I &\frac{V}{2} &. \cr
                    {\! k=2} &. &0 &0 &0 &\frac{V}{2}  &H_0\! + \!2\hbar\omega I &.\cr
                    &. &. &. &. &. &. &.
                   }
\end{align}
which contains the $2 \times 2$ identity matrix $I$ and the $2 \times 2$ zero matrix $0$.
\end{widetext}

Close to resonance we have $\epsilon^{(0)}_{1,k-1}\approx \epsilon^{(0)}_{2,k}$.   The sub matrix corresponding to $u^{(0)}_{1,k-1}$ and $u^{(0)}_{2,k}$ can be read off from the full Floquet Hamiltonian and is found to be
\begin{align}
H'_{\mathrm{F}}=\begin{pmatrix} &\epsilon^{(0)}_{1,k-1} &{\frac{1}{2}{\hbar \eta}} \\
&{\frac{1}{2}{\hbar \eta}} &\epsilon^{(0)}_{2,k} \end{pmatrix}.
\end{align}
When the coupling between these two Floquet states is much stronger than the coupling to other states, which is usually true for two states in resonance, the quasienergies and Floquet states are approximately given by the eigenvalue and eigenvector of the sub matrix. Therefore, at exact resonance where \mbox{$\epsilon^{(0)}_{1,k-1}= \epsilon^{(0)}_{2,k} =\epsilon $}, the quasienergies and Floquet states are 
\begin{align}
\epsilon_{\substack{ 1,{k-1} \\\!\!\!\!\!\!\! 2,k}  }&\approx \epsilon \mp \frac{\hbar \eta}{2}, \nn \\
u_{\substack{1,{k-1} \\\!\!\!\!\!\!\! 2,k }}&\approx \frac{e^{ik\omega t}}{\sqrt{2}}\left(\ket{1}e^{-i \omega t} \mp \ket{2}\right).
\end{align}

Because of the coupling, the two levels are pushed apart by an amount $\delta \epsilon \approx \hbar \eta$. Comparing this result with that for the far off-resonance region of Eqs.~\eqref{offresqe} and \eqref{offresFs}, we observe two important differences: First, the energy shift at resonance is much larger than its far off-resonance value; and second, while for the off-resonance regime the time-dependent part of the wave function is only a small oscillation around the dominant time-independent part, these parts possess comparable amplitudes at resonance. Thus, the wave function oscillates much more strongly when resonance happens and there is little resemblance between the unperturbed wave function and the real one. We conclude that the oscillating field has an important effect on the quasienergy structure and the characteristics of the wave function if there is resonance between any two levels of the unperturbed system. 

For the intermediate values of $\omega_0/\omega$, numerical diagonalization of the Floquet Hamiltonian is needed to obtain the quasienergies. The result rapidly converges with the number $N_f$ of the Floquet modes. In our computation we choose $N_f=20$, which is verified to be sufficiently large. The result is shown in Fig.~\ref{fig2level}(b). The energy shift is indeed maximum at resonance as predicted by the perturbation methods. 

The oscillating potential has another important effect when the parameter $\omega_0$ is varied slowly in an adiabatic process. As can be seen in Fig.~\ref{fig2level}(b), there is an apparent crossing of the unperturbed quasienergy lines at resonance and hence the system started from the branch (1) will continue to move to the same state (1) in the opposite end of the crossing. However, in the real situation when the oscillating potential is  accounted for, the two curves are pushed apart by the coupling and we have an avoided crossing. Thus, the system will move from  state (1) to state (2) in an adiabatic process. Therefore, the dynamics of an adiabatic process passing through a resonance changes completely when the oscillating potential is taken into account.

We must stress that the adiabatic approximation for Floquet states is slightly different from that for the time-independent eigenstates. The mathematical framework of the Floquet adiabatic process for short laser pulses is described in Ref. \cite{Drese99}. In Appendix~\ref{floquetadiabatic} we modify these formulations a little so that it suits a more general problem. A particularly useful result is the Landau-Zener formula for a Floquet adiabatic process which gives the approximate transition probability of a system when it passes through an avoided crossing. For an avoided crossing with asymptotic slope difference $\alpha$ and energy spacing $\delta \epsilon$ as shown in Fig.~\ref{fig2level}(b), the Landau-Zener transition probability is
\beq
P_{\mathrm{LZ}}= \exp \left[-\frac{\pi}{4}\frac{(\delta \epsilon/\hbar)^2}{ \dot{\omega}_0 \tan (\alpha/2) }\right],
\eeq
with $\dot{\omega}_0$ denoting the time derivative of $\omega_0$. When $\alpha$ and $\delta \epsilon$ are known from the energy diagram, the above formula tells us how slow (fast) we need to change the parameter $\omega_0$ to obtain an adiabatic (diabatic) process.

The conclusion that the effect of a weak oscillating field is most important at resonances is also true for a multilevel system. One may use in a similar way the perturbation methods to prove that the energy shift and the oscillation amplitude of a Floquet state $u_{n,k}$ are largest when there is a resonance between the unperturbed level $u^{(0)}_{n,k}$ and another level $u^{(0)}_{m,k-1}$. The split of these levels due to their coupling is  
\beq
\delta \epsilon \approx \left|\bra{n}V\ket{m}\right|. 
\eeq
Thus, a simple way to know whether the exact quasienergies structure differs by a great amount from the approximate unperturbed eigenenergies is to look for resonances in the unperturbed energy diagram and then calculate the coupling strength $ \left|\bra{n}V\ket{m}\right|$ between the resonating levels.

\section{Micromotion effect}\label{micromotioneffect}
We now show how we use the Floquet formalism to evaluate the exact quasienergies of the trapped atom-ion system and from there deduce the effect of the micromotion. As a starting point we assume the atom is fixed at the center of its trap. This corresponds to the situation when the atom is tightly trapped; that is, the atom trapping frequency is very large. We explain later why most of the conclusions about the micromotion effect in this simplified model also hold for a more realistic system.   

In view of the above assumption, the presence of the atom only results in an additional potential in the Hamiltonian of the trapped ion. Therefore, the Hamiltonian of the system given in Eq.~\eqref{atomiontransH} is simplified to
\begin{align}\label{Vatomion}
H(t)= &H_0+H_{\mathrm{mm}}(t),\nn \\
H_0= &\frac{P_i^2}{2m_i} +\frac{1}{2}m_i \omega_0^2 (X_i-d)^2 - \frac{\alpha e^2}{2 X_i^4}, \nn \\ 
 H_{\mathrm{mm}}(t)= & - m_i (\gamma\omega_0)^2 (X_i-d)^2 \cos(2 \omega t)\nn \\
& - \gamma\omega_0 \{X_i - d,P_i\} \sin( \omega t)  .
\end{align}
For the purpose of numerical calculation it is convenient to write the above Hamiltonian in the following dimensionless form:
\begin{align}\label{scaleatomion}
\frac{H(\tau)}{\hbar\omega_0}=&\frac{1}{2}\left(\frac{l_i P_i}{\hbar}\right)^2 + \frac{1}{2} \left(\frac{X_i - d}{l_i} \right)^2 - \frac{1}{2}\frac{(R_i/l_i)^2}{(X_i / l_i)^4} \nn \\
&- \gamma^2 \left(\frac{X_i - d}{l_i} \right)^2 \cos \left(2 \frac{\omega}{\omega_0} \tau\right)  \nn \\
&-\gamma\left\{\frac{X_i-d}{l_i}, \frac{l_i P_i}{\hbar}\right\} \sin \left(\frac{\omega}{\omega_0} \tau\right),
\end{align}
where $R_i=\sqrt{m_i\alpha e^2/\hbar^2}$ is the interaction length redefined for this problem, $l_i=\sqrt{\hbar/m_i \omega_0}$ the harmonic oscillator length, and  $\tau = \omega_0 t$ a new time variable. In short, all lengths are scaled by $l_i$ and energies by $\hbar \omega_0$.

\subsection{Micromotion-induced coupling}\label{micromotioncoupling}
Before we compute the exact quasienergies of the system, it is worth examining whether resonances, as described in Sec.~\ref{resonances}, exist in our system. First we calculate the eigenenergies of the unperturbed Hamiltonian $H_0$. Following Idziaszek \emph{et al.} \cite{Idziaszek07} we compute, at zero trap distance $d=0$, the eigenenergies $E_n(0)$ and eigenstates $\ket{n(0)}$ of $H_0(d=0)$ using the renormalized Numerov method \cite{Johnson77}. The states $\ket{n(0)}$ are then used to diagonalize $H_0$ for other values of the trap distance. To account for the short-range potential, we need to fit the solution to the short-range form \cite{Idziaszek07}
\beq\label{swf}
\Psi(x_i) \propto x_i\sin \left(\frac{R_i}{x_i}+\varphi_{s}\right)
\eeq
for $x_i\ll R_i$. The short-range phase $\varphi_{\mathrm{S}}$ can be related to the \emph{s}-wave scattering length $b$ by
\beq
\cot(\varphi_{s}) = -\frac{b}{R_i}.
\eeq
Once we know the scattering length $b$ and hence the short-range phase, we may find a distance $r_{\mathrm{min}}$ satisfying $a_0~\ll~r_{\mathrm{min}}~\ll~R_i$ such that the wave function given in Eq.~\eqref{swf} vanishes at $r_{\mathrm{min}}$. The boundary condition $\Psi(r_{\mathrm{min}})=0$ is then used to carry out the Numerov computation. In our calculation we assume a single value for the odd and even short-range phases \cite{Idziaszek07}.

\begin{figure}[t]
\centering
\includegraphics[scale=0.35]{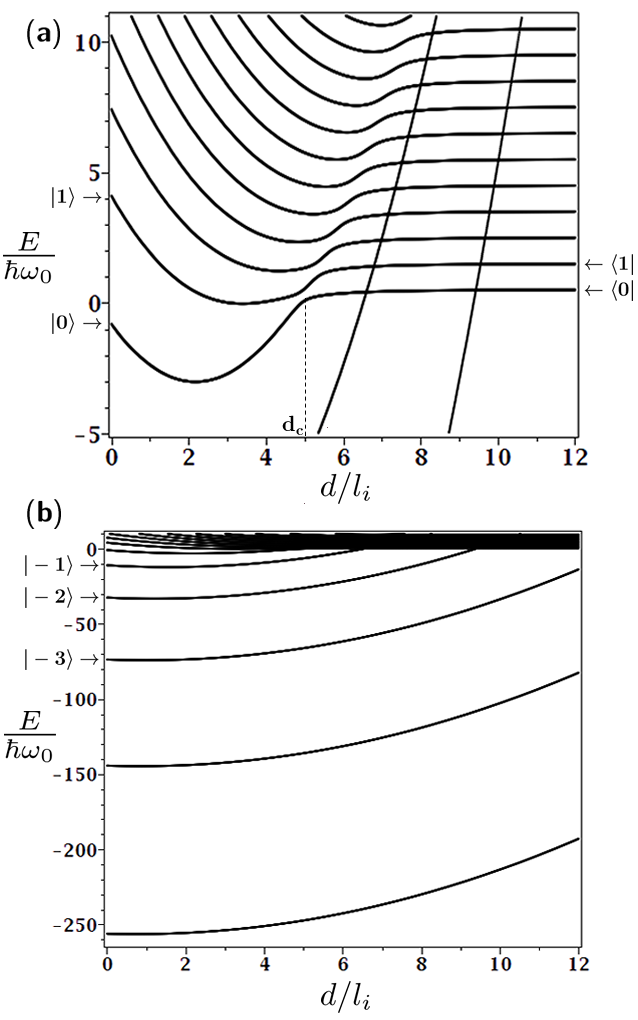}
  \caption{The unperturbed eigenergies of (a) the vibrational levels and (b) the low-lying molecular levels. The energy gaps between adjacent molecular levels are very large compared with those between the vibrational levels which appear as a continuum in the bottom plot.}\label{figee}
\end{figure}

For our numerical calculation we consider the $^{135}\mbox{Ba}^{+}$ ion and $^{87}\mbox{Rb}$ atom system with $R_i= 8927 a_0$. The chosen trapping parameters are $\gamma=1/\sqrt{2}$ ($a=0$), $\omega_0 = 2\pi \times 100$ kHz, $\omega = 2\pi \times 1.27$ MHz; hence, $l_i=516 \, a_0$ and $R_i/l_i\approx 17.2$. The scattering length $b$ is assumed to be $0.9 R_i$, from which we obtain $r_{\mathrm{min}}=0.135 \, l_i$. Although the secular frequency and the micromotion frequency stated above are unrealistically small, we choose these values to obtain plots with better clarity. We repeated our calculation with more realistic values of the frequencies and it did not result in any qualitative change in what we are going to describe about the micromotion effect. More of this is discussed in Sec.~\ref{r1d}.

The eigenenergy diagram of the unperturbed Hamiltonian is shown in Fig.~\ref{figee}. At a large trap distance the spectrum is almost identical to that of a harmonic oscillator as it should be, [see Fig.~\ref{figee}(a)]. At small distances we have a spectrum of vibrational states $(E>0)$ where the ion is localized in the ion trap and molecular states $(E<0)$ where it is bounded to the atom. Note that the energy gaps between adjacent molecular levels shown in Fig.~\ref{figee}(b) are much larger than the ones between the vibrational levels. The molecular states at large trap distances correspond to the situations when we have the traps far apart but the atom and ion are close to each other. These states are, in general, not the initial states in which we prepare our system. The properties of the vibrational as well as molecular states are discussed thoroughly in Refs.~\cite{Doerk10,Idziaszek07}. 

In the proposed experiment for implementing the quantum phase gate \cite{Doerk10}, one starts with the atom and ion cooled to their ground states at a large trap distance and then adiabatically moves the ion trap closer to the atom trap. Thus, we are mainly interested in how the asymptotic ground level changes as the trap distance decreases. From now on when we mention ground level we mean the energy curve that asymptotically coincides with the ground harmonic-oscillator level at a very large trap distance, that is, the curve marked by $\ket{0}$ in Fig.~\ref{figee}(a). The excited energy curves above the ground level are indicated by $\ket{n}$ with positive integers $n$, while the lower molecular energy curves are marked by negative integers. One notices that, when the trap distance is around some characteristic distance $d_c \simeq 5 \,l_i $, the energy of the asymptotic ground state begins to decrease rapidly. This is an indication that the atom-ion interaction is dominant within the range $[0,d_c]$. Furthermore, the energy diagram exhibits an avoided crossing where the ion changes from its  vibrational state to the molecular state around $d_c$. Since the interaction potential must be comparable to the trapping potential at $d_c$, we may roughly estimate this characteristic distance as follows: Suppose the ion is half way between the atom trap and the ion trap; by equating the two potentials we have
\beq\label{dc}
d_c\simeq 2R_i^{1/3}l_i^{2/3} \approx 5.2 \, l_i,
\eeq   
which is close to the value obtained numerically. While this good agreement may not hold for different values of $R_i$ and $l_i$, the above estimation should yield a correct order of magnitude of the ratio $d_c/l_i$.

In order to understand how the oscillating micromotion affects the energy of the ground level we look for resonances between this level and the excited states of the unperturbed system.  In Eq.~\eqref{Vatomion}, the micromotion is represented by two terms with the first oscillating at twice the frequency of the second; hence, there are two types of resonance which we call the $2\omega$ resonance and $\omega$ resonance. 

\begin{figure}[t]
\centering
\includegraphics[scale=0.35]{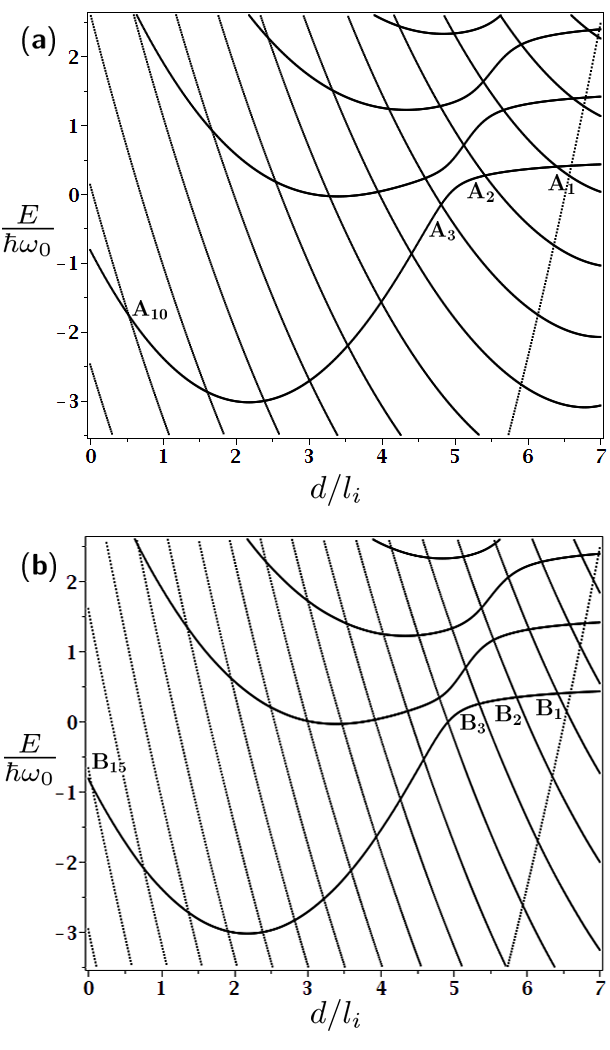}
  \caption{A plot of (a) the eigenenergies $E_n$ together with $E_n -\omega$ shows the $\omega$ resonances and (b) the eigenenergies $E_n$ together with $E_n-2\omega$ shows the $2\omega$ resonances. These resonances appear as crossings. }\label{figres}
\end{figure}

Since in our system $\omega=12.7 \omega_0$, which is not an integral multiple of $\omega_0$, there is no resonance at a large trap distance (we see shortly that resonances at large trap distances are not important anyway). However, as the trap distance is reduced and the energy levels begin to deviate from their asymptotic values, the ground level inevitably comes into resonance with a number of excited levels at different values of the trap distance. To find where the resonances appear, we plot the eigenenergies $E_n$ together with $E_n -\omega$ in Fig.~\ref{figres}(a) and $E_n -2\omega$ in Fig.~\ref{figres}(b), which show that the ground level passes a total number of ten $\omega$-resonances and fifteen $2\omega$-resonances with various excited levels as $d$ goes from 0 to $7 \,l_i$; these resonances appear as crossings and are marked by the letters $\mathrm{A_n}$ and $\mathrm{B_n}$. When the micromotion is included we expect the energy levels to be pushed apart by the micromotion-induced coupling and these apparent crossings to become avoided crossings. 

Next we need to consider the coupling strengths at the resonances. As mentioned in the previous section, the shift in the quasienergy of the ground level can be inferred from the coupling strength $|\!\bra{0}V\ket{n}\!|$, where $\ket{n}$ is the excited state in resonance with the ground level. In Eq.~\eqref{Vatomion}, the micromotion-induced coupling is represented by the operators 
\begin{align}
V_1&=-m_i (\gamma\omega_0)^2 (X_i -d)^2, \nn \\
V_2&=- \gamma\omega_0 \{X_i-d,P_i\}.
\end{align}
Therefore, to study its effect, the quantities $|\!\bra{0}V_{1,2}\ket{n}\!|$ need to be calculated at various trap distances. Since it can be shown that 
\begin{align}\label{couplingv1v2}
\bra{0}V_2\ket{n} = \frac{E_n - E_0}{i \gamma \hbar\omega_0}\bra{0}V_1\ket{n},
\end{align}
the second coupling strength is much larger than the first for sufficiently large $n$ and thus has more important effects. 

\begin{figure*}[t]
\centering
\includegraphics[scale=0.35]{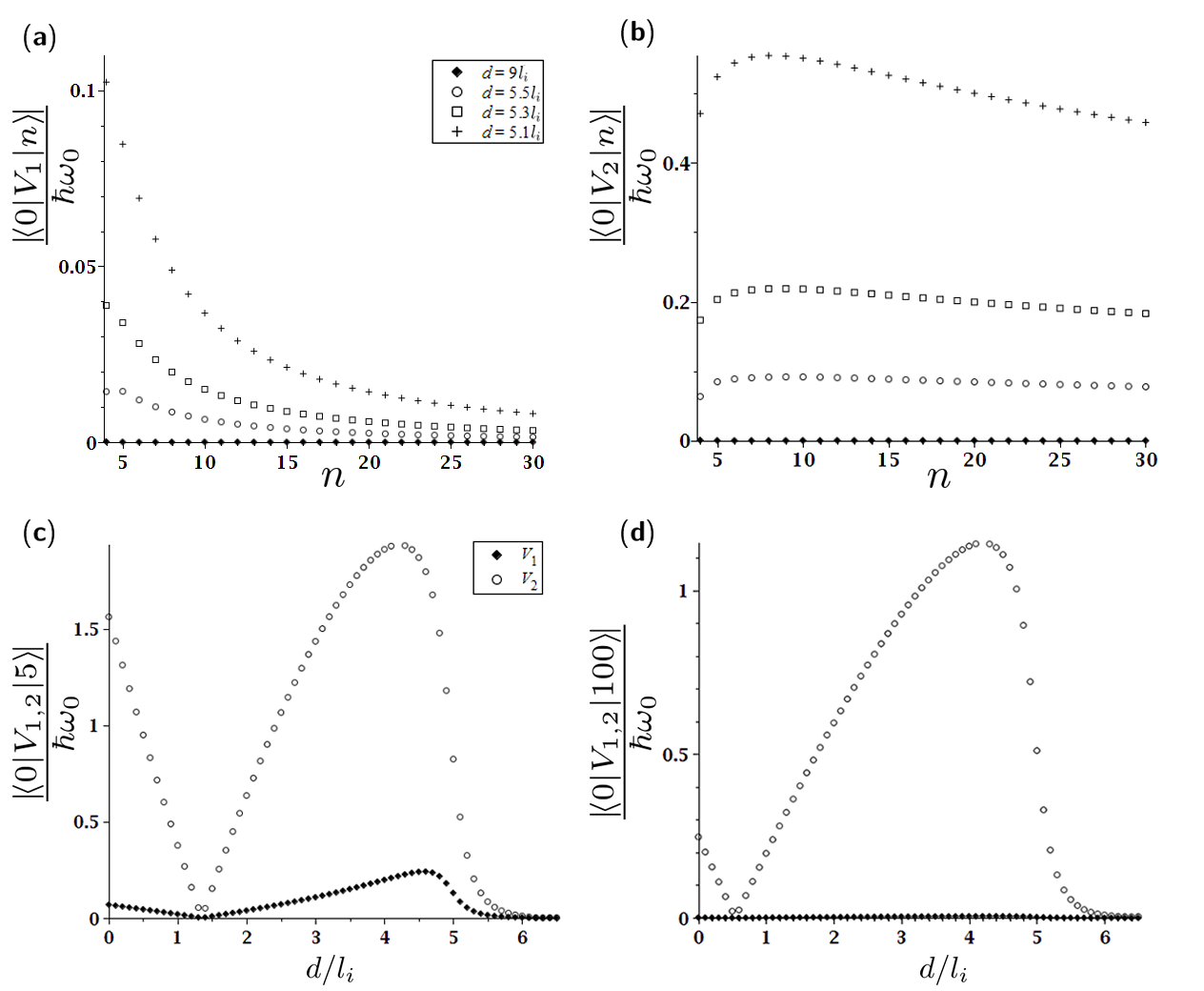}
  \caption{Coupling strengths for (a) $V_1$ and (b) $V_2$ at $d=9\,l_i$ (diamonds), $d=5.5 \,l_i$ (circles), $d=5.3 \,l_i$ (squares), and $d=5.1 \,l_i$ (crosses). The coupling strengths between the ground state and (c) an intermediate excited state $n=5$ and (d) a very high excited state $n=100$ of $V_1$ (diamonds) and $V_2$ (circles) are also plotted as functions of the trap distance.}\label{figcoupling}
\end{figure*}

Figures~\ref{figcoupling}(a) and \ref{figcoupling}(b) show the collective behavior of many coupling strengths, obtained by numerical integration, at four different values of the trap distance $d~=~9\,l_i, 5.5 \,l_i, 5.3\,l_i$ and $d=5.1 \,l_i$. As can be seen in Fig.~\ref{figee}(a), the energy gap between the ground level $\ket{0}$ and the third excited level $\ket{3}$ is at most $11 \omega_0$ at $d=0$. Because this gap is still smaller than the micromotion frequency, we know that there is no resonance between the ground level and the first three excited levels. On the other hand, the energy gap between the ground level and the 26th excited level $\ket{26}$ is too large for resonance. Since the ground level can only come into resonance with those excited levels $\ket{n}$ with $ 4\leq n \leq 25$, only the coupling strengths in that range are important. We see in Figs.~\ref{figcoupling}(a) and \ref{figcoupling}(b)  that when $d=9 \,l_i$, the elements  $|\!\bra{0}V_{1,2}\ket{n}\!|$ are extremely small, which is due to the fact that the atom-ion interaction is very weak at this distance, and hence the unperturbed states are almost identical to the eigenstates of a harmonic oscillator. 
For these states we see immediately $|\!\bra{0_{\mathrm{HO}}}V_{1,2}\ket{n_{\mathrm{HO}}}\!|=0$ for $n>2$. As a result, there is no possible resonance with nonzero coupling strength and we observe in Sec.~\ref{resonances} that the energy shift caused by the micromotion in this case must be very small. In fact, the exact quasienergy spectrum of a single trapped ion was found by Glauber \cite{Glauber07} to be
\begin{align}\label{qe}
\epsilon_n = (n+\frac{1}{2}) \hbar \mu,  
\end{align}
where $\mu$ differs from the secular frequency $\omega_0$ by
\beq
\frac{\mu-\omega_0}{\omega_0} \simeq \left(\frac{\omega_0}{\omega}\right)^2 
\eeq
when $|a|\ll q^2$. This number is very small, so the harmonic approximation works relatively well for trapped ions as long as the external interaction is absent or relatively weak compared with the trapping potential.

The situation changes completely when the ion is subjected to a significant external interaction. At $d=5.5 \,l_i$, where the ion trap just crosses to the strong interaction region, the micromotion-induced coupling begins to increase rapidly and reach a remarkably large value at $d=5.3 \,l_i$. This is due to the now strong enough interaction potential which changes considerably the shape of the unperturbed eigenstates. To see how the coupling strengths change with trap distance we plot a representative one between the ground level and the fifth excited level, that is, $|\!\bra{0}V_{1,2}\ket{5}\!|$, in Fig.~\ref{figcoupling}(c). In addition to the sharp rise at $d\simeq 5.5 \,l_i$, we see that the coupling due to $V_1$ is indeed much weaker than the one due to $V_2$, which is in agreement with Eq.~\eqref{couplingv1v2}.

Since at a small distance the coupling strength of the second term  is comparable to the secular frequency $\omega_0$ while the spacing of the unperturbed energy spectrum is of the order $\omega_0$, we expect the energy diagram to change completely when the oscillating terms of the micromotion are included. More importantly, the fact that the strength of the micromotion-induced coupling increases sharply at a certain value of trap distance means that it can be used to forecast at what characteristic range $d_{\mathrm{mm}}$ the micromotion effect becomes significant and must be taken into account. For our particular system we may predict that the micromotion effect begins to kick in at around $d_{\mathrm{mm}} \simeq 5.5 \,l_i$. 

In the limit of a very large ratio $\omega/\omega_0$, for instance $\omega/\omega_0=100$, resonance is possible only with a large excitation number $n$. As shown in Fig.~\ref{figcoupling}(d), the coupling strength of $V_1$ for large $n$ is essentially zero at all trap distances and therefore can be neglected. On the other hand, Fig.~\ref{figcoupling}(b) shows that the coupling strength of $V_2$ within the range $d\in [0,5.5 \, l_i]$ only decreases slowly with $n$ because it is proportional to the energy difference [see Eq.~\eqref{couplingv1v2}]. Even for $n$ as large as $100$ we have $|\!\bra{0}V_2\ket{100}\!|\simeq 0.1 \hbar\omega_0$ at $d=5.3 \,l_i$ which is not negligible. Moreover, the coupling strength of $V_2$ for the highly excited state $n=100$ exhibits a sudden rise at the same distance $d_{\mathrm{mm}}\simeq 5.5 \,l_i$. We verify that the sharp increase at $d_{\mathrm{mm}}\simeq 5.5\,l_i$ also appears for all $n\in[3,100]$, which implies that the characteristic distance where the micromotion effect becomes important is quite insensitive of the frequency of the trapping potential when the ratio $\omega/\omega_0$ is in the interval $[10,100]$, which contains most of the realistic values for current rf traps.

\subsection{Numerical calculation of exact quasienergies}\label{numerical}
In this section we obtain, by numerical calculation, the quasienergy diagram, from which we then infer the importance of the micromotion effect. There are two widely used methods for computing the quasienergies of a periodic system: Floquet Hamiltonian and time-propagator methods \cite{Chu85}. The Floquet Hamiltonian method is simply the numerical diagonalization of the Floquet Hamiltonian as described in Sec.~\ref{FH}. For our system, it can be expressed in terms of $H_0$ and $H_{\mathrm{mm}}(t)$ in Eq.~\eqref{Vatomion} as
\beq
H_{\mathrm{F}}=H_0 + H_{\mathrm{mm}}(t) -i\hbar \frac{\partial}{\partial t}.
\eeq
We use the Floquet states of the unperturbed Floquet Hamiltonian at zero trap distance $H_{\mathrm{F}}^{(0)}(d=0)$, which are
\beq 
u^{(0)}_{n,k}(0)=\ket{n(0)}e^{i k \omega t},
\eeq
as the basis to evaluate the matrix elements of the Floquet Hamiltonian $H_{\mathrm{F}}$ at an arbitrary trap distance
\begin{align}
&\langle \bra{u^{(0)}_{n,k}(0)}H_{\mathrm{F}}\ket{u^{(0)}_{m,j}(0)}\rangle \nn \\
&=\frac{1}{T}\int_0^T dt e^{-i k\omega t}\bra{n(0)}H_{\mathrm{F}}\ket{m(0)} e^{i j\omega t}.
\end{align}
If one arranges the double indices $\{n,k\}$ such that $n$ runs through all the eigenstates before each change in the Floquet mode $k$, the Floquet Hamiltonian matrix has an infinite block-pentadiagonal structure. With $I$ denoting the identity matrix and $0$ the zero matrix, the Floquet Hamiltonian matrix is
\begin{widetext}
\begin{align} \bordermatrix{ &\ & {-2} &{-1} &{0} &{1} &{j=2}\cr
&. &. &. &. &. &. &. \cr
                  {-2} &. &{H_0\! -\! 2\hbar \omega I} &\frac{i V_2}{2} &\frac{V_1}{2} &0 &0 &.\cr
                   {-1}&. &\frac{-iV_2}{2}  &{H_0\! -\! \hbar\omega I} &\frac{i V_2}{2}  &\frac{V_1}{2} &0 &.\cr
                    {\  \ 0}&. &\frac{V_1}{2} &\frac{-i V_2}{2}  &H_0 &\frac{i V_2}{2}  &\frac{V_1}{2} &.\cr
                   {\  \ 1}&. &0 &\frac{V_1}{2} &\frac{-i V_2}{2}  &H_0\!+\!\hbar\omega I &\frac{iV_2}{2}  &. \cr
                    {\!k= 2}&. &0 &0 &\frac{V_1}{2} &\frac{-i V_2}{2}  &H_0\! +\!2\hbar\omega I &. \cr
                    &. &. &. &. &. &. &. 
                   },
\end{align}
where $H_0$, $V_1$, and $V_2$ are matrices whose elements are 
\begin{align}
\left(H_0\right)_{n,m}=&\left[E_n(0)+ m_i\omega_0^2 d^2/2\right]\delta_{nm} - m_i \omega_0^2 d \bra{n(0)}X_i\ket{m(0)}, \nn \\
\left(V_1\right)_{n,m}=&-m_i (\gamma\omega_0)^2\bra{n(0)}\left(X_i- d\right)^2\ket{m(0)}, \nn \\
\left(iV_2\right)_{n,m}=&\left(\gamma \hbar \omega_0\right)^{-1} \left[E_m(0)-E_n(0)\right]\left(V_1\right)_{n,m}. 
\end{align}
\end{widetext}
Therefore, the Floquet Hamiltonian matrix can be obtained from the matrix elements of $X_i$ and $X_i^2$ at $d=0$. Since we already obtained $E_n(0)$ and $\ket{n(0)}$ by the Numerov method in Sec.~\ref{micromotioncoupling}, the Floquet Hamiltonian matrix elements can be evaluated by numerical integration. 

In practice there are two types of truncation one needs to make. The first comes from the number $N_e$ of the eigenstates $\ket{n(0)}$ and the second comes from the number $N_f$ of the Floquet mode $k$. To generate the basis $u^{(0)}_{n,k}(0)$ we choose all the eigenstates $\ket{n(0)}$ in the energy range $[-5000\, \omega_0, 300 \,\omega_0]$ and $k = -10,\ldots,10$. In total we have $N_e=150$ eigenstates and $N_f =21$ Floquet modes; hence, the linear dimension of our Floquet Hamiltonian matrix is $N_e N_f = 3150$. To obtain a quasienergy diagram, we diagonalize this matrix for trap distances from $0$ to $10 \,l_i$ in step of $\Delta d =0.002 \,l_i$. 

We also use the time-propagator method to calculate the quasienergies of the system. This method is based on Floquet's theorem which states that the time evolution operator satisfying the Schr\"{o}dinger equation 
\beq
i\hbar\frac{\partial}{\partial t} U(t,0) = H(t) U(t,0)
\eeq
has the periodicity property \cite{Salzman74}
\beq
U(t+T,0)=U(t,0)U(T,0),
\eeq
where $U(T,0)$ is a unitary operator whose eigenvalues $\lambda_n$ are related to the quasienergies $\epsilon_n$ by
\beq
\lambda_n =e^{-\frac{i}{\hbar}\epsilon_n T}.
\eeq
So, we need to obtain $U(T,0)$ and diagonalize it to get the quasienergies. First we set $U(0,0)=I$ and compute $U(T,0)$ by using the propagating scheme
\beq
U(t+\Delta t,0) = \exp \left[-i\int_t^{t+\Delta t}dt' H(t')\right]U(t,0) + O(\Delta t^3),
\eeq
which is a second-order method. This simple propagator is the lowest order of the Magnus propagator whose error bound is discussed in Refs.~\cite{Iserles99,Hochbruck03}. For our calculation we use a total number of $N_e=150$ eigenstates in the energy range $[-5000 \,\omega_0, 300 \,\omega_0]$ as a basis to compute the elements of the matrix $H(t)$ and choose $\Delta t= 10^{-3}T$ as the time step. The computation is carried out for trap distances within the interval $[0,10\,l_i]$ in step of $\Delta d =0.002 \,l_i$. In both the Floquet Hamiltonian and the time-propagator methods, we increased $N_e$ up to $253$ and repeated all the calculations to verify that we obtained sufficient accuracies. Here we present the results for $N_e=150$ as the quasienergy plot is clearer with smaller $N_e$.

\begin{figure*}[t]
\centering
\includegraphics[scale=0.56]{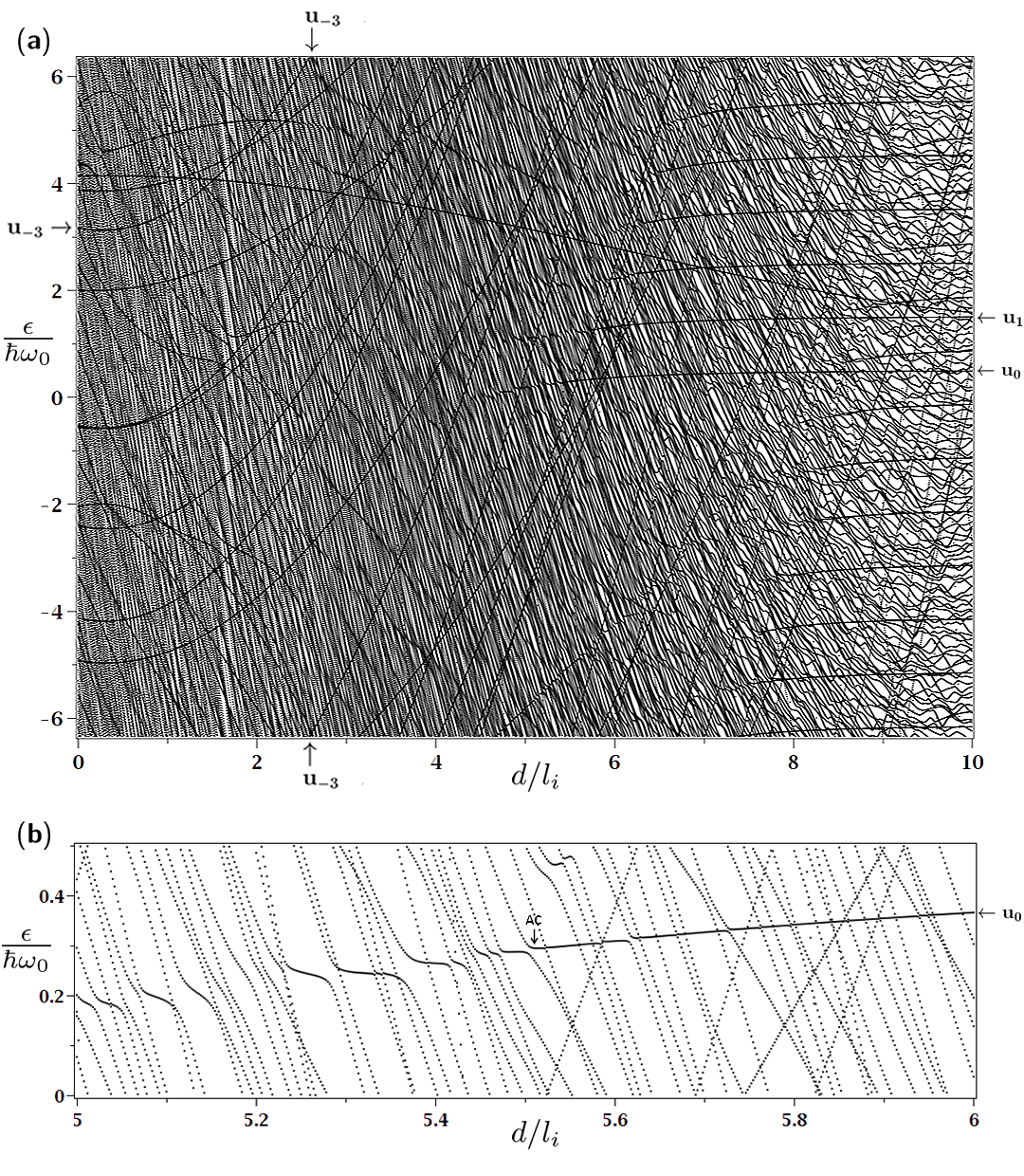}
  \caption{(a) Quasienergies of the trapped atom-ion system plotted in the first zone $[-\hbar\omega/2, \hbar\omega/2]$. While the vibrational energy levels are affected by the micromotion at small trap distances, the energy curves for the low-lying molecular states stay in good agreement with the results obtained by the harmonic approximation. (b) An enlarged plot of the ground level Floquet state. The micromotion-induced avoided crossings appear and the energy curve begins to disintegrate around the characteristics distance $d_{\mathrm{mm}}\simeq 5.7 \,l_i$.}\label{figfullzoneqe}
\end{figure*}

\begin{figure*}[t]
\centering
\includegraphics[scale=0.59]{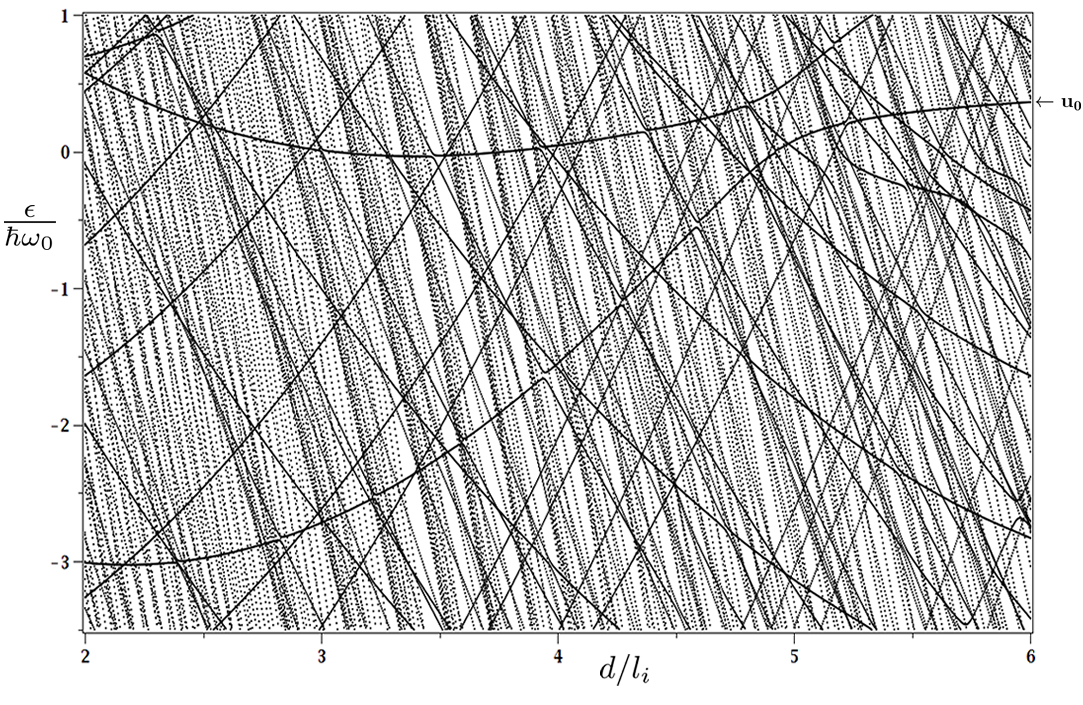}
  \caption{Quasienergy levels obtained when only the first term $V_1$ of the micromotion is included, the energy curve of the ground level is similar to the result obtained by the harmonic approximation except that the new avoided crossings appear at what was the $2\omega$ resonances in the approximate energy diagram shown in Fig.~\ref{figres}.}\label{figqev}
\end{figure*}

We use both the Floquet Hamiltonian and the time-propagator methods to find the quasienergies and observe very good agreement in the results. However, while the Floquet Hamiltonian method is more insightful, the time-propagator method is found to give more accurate results for the quasienergies of the trapped atom-ion system for equal CPU time. In general, it is not convenient to use the Floquet Hamiltonian method when the range of the energy spectrum involved is considerably larger than the frequency of the oscillating field. The quasienergy diagrams shown in Figs.~\ref{figfullzoneqe} and \ref{figqev} are obtained by the time-propagator method. 

As can be seen in Fig.~\ref{figfullzoneqe}, for large trap distances, the quasienergy of the ground Floquet state, marked by $u_0$, agrees quite well with the eigenenergy obtained by the harmonic approximation shown in Fig.~\ref{figee}. The densely distributed curves with steep slopes are quasienergies of the Floquet states that correspond to highly excited unperturbed eigenstates. Since all the quasienergies are returned in the zone $[-\hbar\omega/2, \hbar\omega/2]$, which is an intrinsic property of the Floquet Hamiltonian as well as time-propagator method, these highly excited quasienergy levels are projected down and they appear to cross the ground level. However, at large trap distances, say $d\gtrsim~5.7\,l_i$, the interactions between the ground level and those states are extremely weak and hence the avoided crossings, if they really exist, have very small gaps. Therefore, it is not hard to move the trap distance fast enough to diabatically cross these possible weak avoided crossings so that the system remains in a single Floquet state, which is the ground level in this case.  
 
As the trap distance gets smaller than a characteristic distance $d_{\mathrm{mm}} \simeq 5.7 \,l_i$, the quasienergy begins to differ greatly from the eigenenergy obtained by the harmonic approximation. This is due to the sharp increase in the micromotion-induced coupling discussed in Sec.~\ref{micromotioncoupling}. Within this range, numerous avoided crossings with observable gaps occur in the quasienergy diagram, many of which do not appear in the diagram obtained by the harmonic approximation shown in Fig.~\ref{figee}. The reason is that, at a small trap distance, the coupling of the second term shown in Fig.~\ref{figcoupling} is so strong that it changes the energy diagram completely. 

What does this mean for the implementation of the quantum phase gate? As we slowly change the trap distance passing through $d_{\mathrm{mm}}$, the large number of new avoided crossings of various sizes will lead to branching from the initial ground level to different levels [unless near-perfect experimental control of $d(t)$ is carried out]. The branching of the wave function would result in a poor fidelity $F$ of the phase gate, as we show in Appendix~A that $1-F\propto p_e$, where $p_e$ is the small probability of the ion being in an excited Floquet state at the end of the adiabatic process. Moreover, even if one can control an adiabatic passage of a micromotion-induced avoided crossing, for example the one marked by AC in Fig.~\ref{figfullzoneqe}(b), the duration it takes must be quite long since the energy gap of this avoided crossing is quite small. This long duration would lead to a slow gate speed and hence traversing the micromotion-induced avoided crossings is not desirable. 
\\

The best strategy may be moving the ion trap toward the atom trap until it reaches the characteristics distance $d_{\mathrm{mm}}$ and stop for a while with the hope that the ion picks up enough phase difference for the phase gate. However, it is likely that the energy difference between the triplet and singlet states \cite{Doerk10} at $d_{\mathrm{mm}}$ is very small and a full phase of $\pi$ is not obtainable in a reasonably short time. This energy difference must be determined by experiments because it depends on the short-range phase of the atom-ion interaction potential which is not known. In short, whether stopping at the minimum distance $d_{\mathrm{mm}}$ is good enough cannot be answered by a purely theoretical calculation.

It is not difficult to think of an experiment to detect the characteristic distance $d_{\mathrm{mm}}$ of the micromotion effect. One prepares the ion in the ground Floquet state in the ion trap initially placed very far from the atom trap. The ion trap is then slowly brought closer to the atom trap up to some distance $d_{\mathrm{min}}$ and back to the initial position. If $d_{\mathrm{min}} >d_{\mathrm{mm}}$, one expects the ion to come back to its initial ground state; on the other hand, when $d_{\mathrm{min}} < d_{\mathrm{mm}}$, branching would occur and only a fraction of the ion's wave function is in the ground state. 

As indicated in Fig.~\ref{figfullzoneqe}(a), while the vibrational energy levels are affected by the micromotion at small trap distances, the energy curves for the low-lying molecular states, for instance the ${u_{-3}}$ level, stay in good agreement with the result obtained by the harmonic approximation shown in Fig.~\ref{figee}(b) for all trap distances. Note that the quasienergy curve for the ${u_{-3}}$ level goes out at the top and comes back at the bottom of the picture. The fact that the molecular Floquet states have smooth energy curves without any avoided crossing of observable size means that the micromotion does not greatly affect these states. The reason is that the energy gap between a low unperturbed molecular state and other states, as shown in Fig.~\ref{figee}(b), are much larger than the micromotion frequency $\omega$ and hence the oscillating terms of the micromotion cannot couple the unperturbed molecular state to other states. In other words, the unperturbed molecular state does not come into resonance when the trap distance is changed from $10 \,l_i$ to $0$, and as a result the oscillating terms have little effect on the energy structure of these states. Consequently, the wave function of these Floquet molecular states must consist of a dominant time-independent part which is just the unperturbed wave function and a small time-dependent part that oscillates at the frequency $\omega$. As we show in the next section, the fact that these molecular states are almost unaffected by the micromotion offers a way to implement the quantum phase gate despite the difficulties caused by the micromotion that we discussed in the preceding paragraphs. 

To compare the effect of the two oscillating terms of the micromotion, we calculate the quasienergies of the system when only the first term $V_1$ is included. The result is shown in Fig.~\ref{figqev}; the energy curve of the ground level is similar to the result obtained by the harmonic approximation except that new avoided crossings appear at what was the $2\omega$ resonances in the approximate energy diagram shown in Fig.~\ref{figres}(b). This structure can be explained by the lowest-order perturbation theory described in Sec.~\ref{resonances}. It confirms the observation we made in Sec.~\ref{micromotioncoupling} that the first term only has a weak effect. Although they seem to be insignificant, the new avoided crossings lead to a completely different dynamics of the system in an adiabatic process. This means an oscillating term, even if relatively weak, must still be taken into account if it results in resonances between the unperturbed energy levels. 

We see in Figs.~\ref{figfullzoneqe}(a) and \ref{figqev} that most of the dramatic change of the energy structure is caused by the second term $V_2$ of the micromotion. This coupling is so strong that perturbation methods are no longer good enough to account for its influence. Because $V_2$ contains the momentum operator $P$ which is related to the velocity of the micromotion, the intuition that the micromotion must be important because its classical velocity is large is correct after all. In the limit of very large ratio $\omega/\omega_0$, we may neglect the first term $V_1$ in the micromotion Hamiltonian for the reasons explained in the last paragraph of Sec.~\ref{micromotioncoupling}. In the same paragraph we also showed why the characteristic distance $d_{\mathrm{mm}}$ should not change when the ratio $\omega/\omega_0$ increases up to $100$. Since $V_1$ can be ignored when $\omega/\omega_0\simeq 100$ , numerical computation of the quasienergies is considerably simpler in this limit.

Last but not least, we must stress an important point regarding the numerical calculation of the quasienergies of a trapped atom-ion system. One may notice that the Floquet Hamiltonian and time-propagator method can be used with the original Hamiltonian given in Eq.~\eqref{atomionH}. The reason we always work in the transformed picture is that the Cook-Shankland transformation reduces the magnitude of the time-dependent terms in the Hamiltonian by a factor of $\omega_0/\omega$ and hence much faster numerical convergence is achieved. In fact, given the same size of the Floquet Hamiltonian matrix, the accuracy  of the quasienergies obtained when working with the original Hamiltonian is much poorer than that of the quasienergies obtained in the transformed picture.
 
\subsection{A scheme to bypass the micromotion effect}\label{bypass}
We already showed that an adiabatic process using the ground level may not be practically possible because of the micromotion effect. However, the fact that the low-lying molecular states are almost unaffected by the micromotion means that we can bypass the micromotion effect by making a transition from the ground level to one of the molecular levels at a distance larger than $d_{\mathrm{mm}}$ before the micromotion becomes important. More specifically, we may prepare the ion in the ground Floquet state at a large trap distance and then move it adiabatically toward the atom trap. When the trap distance reaches some value $d_1$ which is slightly larger than $d_{\mathrm{mm}}$, we make the transition from the ground level $u_0$ to a molecular level lying below whose wave function has a sufficient overlap with $u_0$. This transition can be realized by applying an electromagnetic field $\vec{E}(t)$. When the system is in this Floquet molecular state it will be ``protected" from the micromotion and we may continue the adiabatic process up until $d=0$. This modified adiabatic process can be used to implement the quantum controlled phase gate provided that the transition strengths for the triplet and singlet states are equal at $d_1$. 

In principle, we need to investigate the transition between two Floquet states: $u_0$ and a Floquet molecular state, say the one lying two levels below the ground level, $u_{-3}$. Nevertheless, when $d>d_1$ we know that the micromotion has little effect and the dominant parts of these Floquet states are given by the unperturbed eigenstates $\ket{0}$ and $\ket{-3}$ shown in Fig.~\ref{figee}; hence, it is possible to use the states $\ket{0}$ and $\ket{-3}$ to study the transition between the exact Floquet states without encountering too much error. 

Suppose the one-dimensional system described in Eq.~\eqref{atomiontransH} is subjected to an electromagnetic field, 
\beq
\vec{E}(t)=E_0 \cos(\omega_f t)\vec{e}_x.
\eeq
The coupling energy resulting from the interaction of this field with the atom's induced dipole $p_a = 4\pi \epsilon_0\alpha e/x_i^2$ and the ion's charge $e$ is 
\beq
V_{\mathrm{cp}}= \frac{4\pi \epsilon_0\alpha e}{X_i^2}E(t) - eX_i E(t).
\eeq  

If we choose the distance at transition to be $d_1=6 \,l_i$, the frequency $\omega_f$ of the electromagnetic field must satisfy the resonance condition
\beq
\hbar\omega_f =E_0(6\,l_i)-E_{-3}(6\,l_i),
\eeq
and a quick glance at Fig.~\ref{figee}(b) reveals that
\beq
\omega_f\approx 62 \, \omega_0 = 2 \pi \times 6.2 \mbox{MHz}.
\eeq
We need to evaluate the matrix element $|\!\bra{0}V_{\mathrm{cp}}\ket{-3}\!|$ to know whether the coupling is strong enough to make a transition in practice. At $d_1$ the first term in the potential $V_{\mathrm{cp}}$ is negligible compared with the second term; thus, we only need to compute the element $|\!\bra{0}X_i\ket{-3}\!|$. This gives
\beq
|\!\bra{0}X_i\ket{-3}\!|\approx 1.3 \times 10^{-4} \,l_i. 
\eeq
The Rabi frequency of the transition is 
\beq
\Omega = \frac{eE_0}{\hbar}|\!\bra{0}X_i\ket{-3}\!|,    
\eeq
and for an electric field strength of $2 \,\mbox{Vcm}^{-1}$ we obtain a Rabi frequency of around $1 \, \mbox{MHz}$ which corresponds to a sufficiently strong transition. The numerical values required for the frequency $\omega_f$ and the field strength are both reasonable; thus, it is practically possible to carry out a transition from $u_0$ to $u_{-3}$ at $d_1$. A transition from a vibrational state to a molecular state may also be realized by Raman transitions. 

Another interesting application of the scheme above is the creation of an atom-ion macro molecule. At the end of the process when the atom and ion are in the state $u_{-3}$ at $d=0$, it is possible to make a transition to a much lower molecular state. After that we may turn off the atom trap and what is left is a bounded atom-ion molecule stored in the ion trap. The same procedure between the resulting ion and another atom can be repeated to create a larger molecule as long as the stability condition for the ion trap is not violated. Such an atom-ion macromolecule may well be of interest in its own right.

\subsection{Excess micromotion}\label{excesmicro}

In practice, there are two types of micromotion in an ion trap. The one we described so far is the intrinsic micromotion which comes from the driving potential. The other type of micromotion is the excess micromotion which comes from imperfections in experimental setups. In Ref.~\cite{Berkeland98}, Berkeland \emph{et al.} give a detailed description of the excess micromotion and its unwanted effects in high-resolution spectroscopy of an ion in a linear Paul trap. Ideally, the excess micromotion must be eliminated. Here we investigate a realistic situation where a small amount of the excess micromotion is left in the motion of a trapped ion. 

Basically, the excess micromotion is caused by a uniform static electric field $\vec{E}_{\mathrm{dc}}$ or a phase difference $\varphi_{\mathrm{ac}}$ between the ac potentials of the ion trap's electrodes. This phase shift results in an oscillating electric field $\vec{E}_{\mathrm{ac}} \sin{(\omega t)}$, where $\omega$ is the frequency of the driving potential. With $R_t$ denoting the radius of the trap's cylindrical electrodes, the amplitude $E_{\mathrm{ac}}$ is related to the phase difference $\varphi_{\mathrm{ac}}$ by \cite{Berkeland98} 
\beq\label{eac}
E_{ac}\simeq \frac{m_i}{e} q\omega^2 R_t \varphi_{ac}.
\eeq

Suppose $\vec{E}_{\mathrm{dc}}$ and $\vec{E}_{\mathrm{ac}}$ points along the $x$ axis; the Schr\"odinger equation for the motion along the $x$ axis of a single trapped ion is now
\begin{align}\label{eSe}
i\hbar\frac{\partial}{\partial t} \Psi(x,t)= &\Big\{-\frac{\hbar^2}{2m_i}\frac{\partial^2}{\partial x^2} +\frac{1}{8}m_i \omega^2 x^2[a+2q \cos(\omega t)]  \nn \\
& - eE_{\mathrm{dc}}x - eE_{\mathrm{ac}}x\sin (\omega t)\Big\}\Psi(x,t).
\end{align}
The term proportional to $x^2$ in the potential describes the driving potential in the ideal case when there is no excess micromotion. This potential leads to the intrinsic micromotion whose effect has been discussed in previous sections. The other terms which are linear in $x$ describe the excess micromotion.

To investigate the effect of the excess micromotion, we first follow Cook \emph{et al.} \cite{Cook85} and ignore all the time-independent operators in the Schr\"odinger equation. The solution is then, obviously,
\begin{align}
\Psi(x,t)=\Psi(x,0)  \exp \Big\{&-\frac{i}{\hbar}\Big[\frac{q}{4}m_i\omega x^2 \sin (\omega t) \nn \\
&+\frac{eE_{\mathrm{ac}}}{\omega}x\cos (\omega t)\Big]\Big\}.
\end{align}
The phase factor here reduces to the Cook-Shankland phase in Eq.~\eqref{phase} when $E_{ac}$ vanishes. Again, one may argue that the main effect of the time-dependent terms in the potential is to introduce this oscillating phase to the wave function. Therefore, we are motivated to write the wave function in Eq.~\eqref{eSe} as
\begin{align}
\Psi(x,t)= \exp \Big\{&-\frac{i}{\hbar}\Big[\frac{q}{4}m_i\omega x^2 \sin (\omega t) \nn \\
&+\frac{eE_{\mathrm{ac}}}{\omega}x\cos (\omega t)\Big]\Big\} \Phi(x,t)
\end{align}
with the expectation that the time-dependent part of $\Phi(x,t)$ is sufficiently small. A substitution of the above expression into Eq.~\eqref{eSe} yields the following equation for $\Phi(x,t)$:
\begin{widetext}
\begin{align}
i\hbar\frac{\partial}{\partial t} \Phi(x,t) = \Big\{&-\frac{\hbar^2}{2m_i}\frac{\partial^2}{\partial x^2} +\frac{1}{2}m \omega_0^2 x^2 -eE_{dc}x- m_i (\gamma\omega_0)^2 x^2 \cos(2\omega t) +2 i\hbar \gamma\omega_0 \left(x\frac{\partial}{\partial x}+\frac{1}{2}\right)\sin(\omega t)\nn \\
& +\frac{\gamma \omega_0}{\omega}eE_{ac}x\sin (2\omega t)+i\hbar\frac{eE_{ac}}{m_i\omega}  \cos(\omega t) \frac{\partial}{\partial x}+\frac{1}{2m_i}\left(\frac{eE_{\mathrm{ac}}}{\omega}\right)^2 \left[\cos (\omega t)\right]^2 \Big\}\Phi(x,t).
\end{align}
\end{widetext}
The last term gives rise to a time-dependent global phase in the wave function with no physical meaning and hence can be discarded. Consequently, the Hamiltonian of a single trapped ion in the transformed picture is 
\beq\label{eH}
H(t)=H_0+H_{\mathrm{mm}}(t)+H_{\mathrm{ex}}(t),
\eeq
where  
\beq
H_0=\frac{P^2}{2m_i} + \frac{1}{2} m_i \omega_0^2 X^2 -eE_{\mathrm{dc}}X
\eeq
is the unperturbed Hamiltonian, $H_{\mathrm{mm}}(t)$ is the Hamiltonian of the intrinsic micromotion which has the same form as in Eq.~\eqref{im}, and $H_{\mathrm{ex}}(t)$, the Hamiltonian of the excess micromotion, is 
\beq
H_{\mathrm{ex}}(t)=\frac{\gamma\omega_0}{\omega} eE_{\mathrm{ac}} X \sin (2\omega t)-eE_{\mathrm{ac}}\frac{P}{m_i\omega}\cos (\omega t).
\eeq

In Appendix B we demonstrate that the addition of the excess micromotion does not lead to any change in the quasienergies of a single trapped ion. In other words, the ion possesses a harmonic oscillator-like  quasienergy spectrum as given in Eq.~\eqref{qe}, where $\mu$ is a function of $\omega,a,q$ and does not depend on $E_{\mathrm{dc}}$ and $E_{\mathrm{ac}}$. However, the excess micromotion does lead to significant changes in the structure of the Floquet states.

Now let us work out the effect of the excess micromotion for the one-dimensional trapped atom-ion system in which the atom is fixed at the center of the atom trap as described in Fig.~\ref{figtrapatomion}. In this case, the presence of the atom results only in an additional interaction potential in the Hamiltonian given in Eq.~\eqref{eH}. Let us choose the location of the atom as the coordinate origin, then the Hamiltonian of the system is 
\begin{align}
H(t)=&H_0+H_{\mathrm{mm}}(t)+H_{\mathrm{ex}}(t),\nn \\
H_0=&\frac{P_i^2}{2m_i} +\frac{1}{2}m_i \omega_0^2 (X_i-d)^2\! - eE_{\mathrm{dc}}(X_i-d)\!-\frac{\alpha e^2}{2X_i^4}, \nn \\ 
 H_{\mathrm{mm}}(t)\!=& - m_i (\gamma\omega_0)^2 (X_i-d)^2 \cos(2 \omega t)\nn \\
&- \gamma\omega_0 \{X_i - d,P_i\} \sin( \omega t)  , \nn \\
H_{\mathrm{ex}}(t)=&\frac{\gamma \omega_0}{\omega} eE_{\mathrm{ac}}(X_i\!-\!d)\! \sin (2\omega t)\!- \!eE_{\mathrm{ac}}\frac{P_i}{m_i\omega}\!\cos(\omega t).
\end{align}
Thus, the contribution of the field $E_{\mathrm{dc}}$ is to change the structure of the unperturbed eigenstates. Apart from an additive constant, the unperturbed Hamiltonian $H_0$ can be expressed as
\begin{align}
H_0=\frac{P_i^2}{2m_i} +\frac{1}{2}m_i \omega_0^2 \left(X_i-d-\frac{eE_{\mathrm{dc}}}{m_i\omega_0^2}\right)^2-\frac{\alpha e^2}{2 X_i^4}.
\end{align}
Thus the effect of the dc electric field amounts to shifting the trap distance by 
\beq\label{le}
\delta d = \frac{eE_{\mathrm{dc}}}{m_i \omega_0^2}.
\eeq
For a small value of $E_{\mathrm{dc}}$, say, $0.01 \, \mbox{Vm}^{-1}$, with the values of $m_i$ and $\omega_0$ as given in the numerical calculation carried out in Sec.~\ref{micromotioneffect}, the distance shift is $\delta d \approx 0.7\, l_i$. Here the distance shift is already significant and it can be even larger when either the electric field $E_{\mathrm{dc}}$ increases or the secular frequency $\omega_0$ decreases. Thus, care must be exercised when such a uniform electric field exists in the trap.

It is advantageous to make the change of variable $d'=~d+\delta d$ to bring the form of $H_0$ to that of the unperturbed Hamiltonian we considered in Eq.~\eqref{Vatomion}. Then the eigenenergies and eigenstates of $H_0$ at various values of $d'$ can be readily taken from the calculation done in Sec.~\ref{micromotioncoupling}. With this transformation, the micromotion Hamiltonian can be written as the sum $H'_{\mathrm{mm}}(t)+H_{\mathrm{ac}}(t)+~H_{\mathrm{dc}}(t)$, where the forms of $H'_{\mathrm{mm}}(t)$ and $H_{\mathrm{ac}}(t)$ are, respectively, those of $H_{\mathrm{mm}}(t)$ and $H_{\mathrm{ex}}(t)$ with $d$ replaced with $d'$, and $H_{\mathrm{dc}}(t)$ is 
\begin{align}
H_{\mathrm{dc}}(t)=& -2 m_i (\gamma\omega_0)^2 \delta d (X_i-d') \cos(2 \omega t) \nn \\
&-2 \gamma \omega_0 (\delta d) P_i \sin( \omega t).
\end{align}

So, by the change of variable we may now view the unperturbed system as unaffected while the dc field effectively results in two additional oscillating terms in the Hamiltonian. The fact that a dc electric field leads to not only a displacement of the ion but also an additional micromotion is not a surprise because an off-centered ion is subjected to a stronger force of the driving rf field.

To compare the magnitude of the excess micromotion caused by the ac field $H_{\mathrm{ac}}(t)$ and the intrinsic micromotion $H'_{\mathrm{mm}}(t)$, we introduce the length scale $l_{\mathrm{ac}}$ associated with the force $eE_{\mathrm{ac}}$ by
\beq
eE_{\mathrm{ac}}=m_i\omega \omega_0 l_{\mathrm{ac}}
\eeq
and write the Hamiltonians of the excess micromotion in the following dimensionless forms:
\begin{align}
\frac{H_{\mathrm{ac}}(t)}{\hbar \omega_0}=\frac{l_{\mathrm{ac}}}{l_i}\left[\gamma \frac{X_i-d'}{l_i}\sin (2\omega t)-\frac{l_iP_i}{\hbar}\cos(\omega t)\right]
\end{align}
and
\beq
\frac{H_{\mathrm{dc}}(t)}{\hbar \omega_0}=-2\gamma\frac{\delta d}{l_i}\left[\gamma \frac{X_i-d'}{l_i}\cos (2\omega t)+\frac{l_iP_i}{\hbar}\sin(\omega t)\right].
\eeq
By a comparision with the dimensionless form of $H'_{\mathrm{mm}}(t)$ given in Eq.~\eqref{scaleatomion} (with $d$ replaced by $d'$), we see that the relative strength of the excess micromotion is described by the factors $l_{\mathrm{ac}}/l_i$ and $\delta d/l_i$. From Eqs.~\eqref{eac} and \eqref{le} we obtain
$
l_{\mathrm{ac}}\simeq R_t\varphi_{\mathrm{ac}}.
$
For a millimeter-sized trap with electrode radius $R_t\simeq 1$ mm, a phase shift as small as $\varphi_{\mathrm{ac}} \simeq 10^{-2}$ degrees yields $l_{\mathrm{ac}}/l_i \simeq 6$. Thus, the effect of the ac excess micromotion is comparable to that of the intrinsic micromotion even when the phase shift is very small. To avoid further complications due to the excess micromotion we have to keep this phase shift and the electric field $E_{\mathrm{dc}}$ as small as possible.  

\begin{figure}[t]
\centering
\includegraphics[scale=0.42]{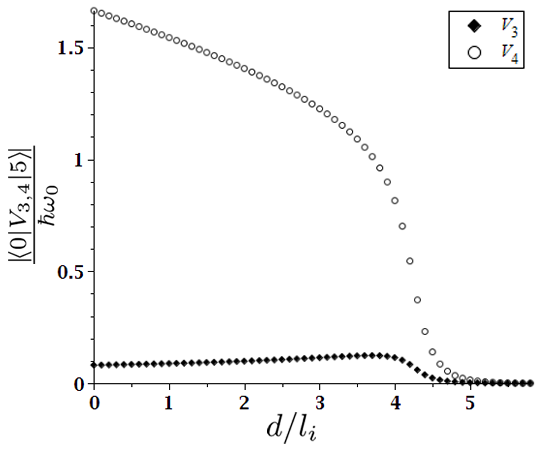}
  \caption{A plot of the coupling strength  $|\!\bra{0}V_{3,4}\ket{5}\!|$ as a function of trap distance.}\label{figexcess}
\end{figure}

In circumstances when $\delta d/l_i$ and $l_{\mathrm{ac}}/l_i$ are not negligible, the numerical methods described in Sec.~\ref{numerical} can be readily applied to obtain the quasienergy diagram of the system. However, for a qualitative picture, it suffices to compute the coupling strength as described in Sec.~\ref{micromotioncoupling}. When the excess micromotion is considered, we need to calculate six instead of two coupling strengths. Figure~\ref{figexcess} shows the representative coupling strength $|\!\bra{0}V_{3,4}\ket{5}\!|$, plotted against the original trap distance $d$, for the operators
\begin{align}
V_3&=\frac{\gamma\omega_0}{\omega} eE_{\mathrm{ac}}(X_i-d), \nn \\
V_4&=-eE_{\mathrm{ac}}\frac{P_i}{m_i\omega},
\end{align}
of the ac excess micromotion Hamiltonian $H_{\mathrm{ac}}(t)$; the two operators of $H_{\mathrm{dc}}(t)$ only differ from $V_{3,4}$ by the factor $2\gamma\delta d/l_{\mathrm{ac}}$. There is a sharp rise at around $d_{\mathrm{mm}} \simeq 4.8 \,l_i$, which differs from the corresponding value $d_{\mathrm{mm}} \simeq 5.3 \,l_i$ obtained in Sec.~\ref{micromotioneffect}A by $\delta d = 0.7 \,l_i$, as it should be. We observe that the coupling strengths, within the range of resonance, of all the operators in $H'_{\mathrm{mm}}(t)$, $H_{\mathrm{ac}}(t)$, and $H_{\mathrm{dc}}(t)$ also exhibit a sudden increase at $d_{\mathrm{mm}} \simeq 4.8 \,l_i$.

We may now describe the key behaviors of the quasienergy diagram. At large trap distances, that is, $d >~d_{\mathrm{mm}}$, the interaction potential can be neglected and the quasienergy of the system is essentially the energy of a single trapped ion. Recall that the excess micromotion does not change the quasienergy of a single trapped ion, the asymptotic quasienergies of the system must still be given by Eq.~\eqref{qe}. When the trap distance decreases we expect resonances to happen and micromotion-induced avoided crossings to occur in the quasienergy diagram. As a result, the qualitative behavior of the quasienergy curves shown in Fig.~\ref{figfullzoneqe} still holds when excess micromotion of reasonable magnitudes exists in the system. The only difference expected is a shift of $\delta d$ in the micromotion characteristic distance $d_{\mathrm{mm}}$.

\subsection{A more realistic 1D model}\label{r1d}
The simplified model of a fixed atom interacting with a trapped ion considered in the previous sections helps us understand the quasienergy structure of a more realistic system. In fact, all of the conclusions we obtained from the simplified model are still qualitatively true when we allow the trapped atom to move. The Hamiltonian of this system is then given in Eq.~\eqref{atomiontransH}. Let us consider the case when $\omega_a=\omega_0$. By switching the coordinates to those of the relative motion 
\beq
X=X_i-X_a,
\eeq
and the center-of-mass motion  
\beq
X_{\mathrm{cm}}=\frac{m_a X_a+m_i \left(X_i-d\right)}{m_a+m_i},  
\eeq
we may write the Hamiltonian of the system as
\begin{widetext}
\begin{align}
H(t)&=H_{\mathrm{cm}}(t)+H_{\mathrm{rel}}(t)+H_1(t),\nn \\
H_{\mathrm{cm}}(t)&=\frac{P^2_{\mathrm{cm}}}{2M}+\frac{1}{2}M\omega_0^2 X_{\mathrm{cm}}^2 - \frac{m_i}{M}\left[ M (\gamma\omega_0)^2 X_{\mathrm{cm}}^2 \cos(2\omega t) + \gamma\omega_0 \{X_{\mathrm{cm}},P_{\mathrm{cm}}\} \sin( \omega t) \right], \nn \\
H_{\mathrm{rel}}(t)&=\frac{P^2}{2m}+\frac{1}{2}m\omega_0^2 \left(X-d\right)^2 - \frac{\alpha e^2}{2X^4} -\frac{m}{m_i}\left[ m (\gamma\omega_0)^2 \left(X-d\right)^2 \cos(2\omega t) +\gamma\omega_0 \{X-d,P\} \sin( \omega t) \right], \nn \\
H_1(t)&=-2 m (\gamma \omega_0)^2 X_{\mathrm{cm}}\left(X-d\right)\cos (2\omega t)-2 \gamma \omega_0 \left[ X_{\mathrm{cm}}P+\frac{m}{M}\left(X-d\right)P_{\mathrm{cm}}\right]\sin (\omega t),
\end{align}
with the total mass $M=m_a+m_i$ and the reduced mass $m=m_a m_i/(m_a+m_i)$.
\end{widetext}
Without the micromotion, the Hamiltonians for the center-of-mass mode and relative mode are decoupled. It is the term $H_1(t)$ caused by the micromotion that couples these modes. It is clear that the Hamiltonian of the relative motion $H_{\mathrm{rel}}(t)$ is, apart from some multiplicative constants, the Hamiltonian of the simplified model we considered in Sec.~\ref{micromotioncoupling}. This was our motivation for the simplification in the first place. 

The unperturbed eigenstates of the system are the product states 
\beq
\ket{n_{\mathrm{cm}},n}=\ket{n_{\mathrm{cm}}}\ket{n},
\eeq
and the unperturbed eigenenergies are the sum
\beq
E_{n_{\mathrm{cm}},n}=\left(n_{\mathrm{cm}}+\frac{1}{2}\right)\hbar \omega_0+ E_{n}.
\eeq
where $\ket{n}$ and $E_{n}$ are the unperturbed eigenstates and eigenenergies of the relative Hamiltonian as discussed in Sec.~\ref{micromotioncoupling}. The only difference is the replacement of the ion mass $m_i$ by the reduced mass $m$ and hence there is no qualitative change in the eigenenergy diagram as shown in Fig.~\ref{figee}.

Now let us analyze how the quasienergy of the ground Floquet state looks like when the trap distance is varied in an adiabatic process. The asymptotic quasienergies of the system are the sum of the eigenenergies of the trapped atom,
\beq
E_a=\left(n_{a}+\frac{1}{2}\right)\hbar \omega_0,
\eeq
and the quasienergy of a single trapped ion given in Eq.~\eqref{qe}.
Furthermore, the ground Floquet state of the system is the product $\ket{0_a}\ket{u_0(t)}$ of the ground harmonic oscillator state of the atom and the ground Floquet state of the ion. The asymptotic form of the ground Floquet state and its quasienergy are well approximated by the unperturbed eigenstates $\ket{n_{\mathrm{cm}}=0,n=0}$ and their eigenvalues $E_{0,0}$. 

When we decrease the trap distance, the micromotion will become important when the ground unperturbed state $\ket{0,0}$ comes into resonance with an excited state $\ket{n_{\mathrm{cm}},n}$, that is,
\beq
E_{n_{\mathrm{cm}},n}-E_{0,0}=\hbar \omega.
\eeq
If the coupling strength $|\!\bra{0,0}V\ket{n_{\mathrm{cm}},n}\!|$ of the oscillating terms in $H(t)$ is sufficiently large, micromotion-induced avoided crossings of observable size will occur along the quasienergy curve of the ground Floquet state. 

We need to compute the coupling strengths to understand the impact of the micromotion. For the two oscillating terms in $H_{\mathrm{cm}}(t)$ they are, apart from the factor $m_i/M$ of order one,
\begin{align}
M(\gamma\omega_0)^2&\left|\bra{0,0} X_{\mathrm{cm}}^2 \ket{n_{\mathrm{cm}},n}\right| \nn \\
&= \frac{\gamma^2}{2} \hbar \omega_0 \left(\sqrt{2}\delta_{n_{\mathrm{cm}},2}+\delta_{n_{\mathrm{cm}},0}\right)\delta_{n,0}, \end{align}
and
\begin{align}
\gamma \omega_0&\left|\bra{0,0}\{X_{\mathrm{cm}},P_{\mathrm{cm}}\}\ket{n_{\mathrm{cm}},n}\right|\nn \\
&=\frac{\gamma}{2}n_{\mathrm{cm}}\hbar\omega_0  \left(\sqrt{2}\delta_{n_{\mathrm{cm}},2}+\delta_{n_{\mathrm{cm}},0}\right)\delta_{n,0}.
\end{align}
These couplings are nonvanishing only if $n=0$; that is, resonances must happen in the subspace of the center-of-mass unperturbed eigenstates for them to be of any importance. Moreover, as long as $\omega\gtrsim 10 \,\omega_0$ these resonances are only possible if $n_{\mathrm{cm}}>2$, which implies that the coupling strengths described above always vanish at resonances. Hence, we conclude that the two oscillating terms in $H_{\mathrm{cm}}(t)$  have negligible effect. This is not a surprise when we notice that $H_{\mathrm{cm}}(t)$ is just the Hamiltonian of a single trapped ion for which the micromotion is known to have little effect in the energy structure of the first few states.

Similarly, for the two oscillating terms in $H_{\mathrm{rel}}(t)$, the coupling strengths are nonvanishing only for resonances in the subspace of the relative-mode unperturbed eigenstates. Therefore, these couplings reduce to the ones we discussed in Sec.~\ref{micromotioncoupling}. By repeating the numerical calculation shown in that section with $m_i$ replaced by $m$, we find that these coupling strengths are significant only when $d \lesssim 4.8 \, l_{\mathrm{rel}}$ with $l_{\mathrm{rel}}=\sqrt{\frac{\hbar}{m \omega_0}}\approx 824 \,a_0$.  

Finally, the coupling strengths for the oscillating terms in $H_1(t)$ are nonvanishing only if $n_{\mathrm{cm}}=1$, which corresponds to the the following sideband resonance in the subspace of the unperturbed eigenstates of the relative mode
\beq
E_{n}-E_{0}=\hbar\left(\omega-\omega_0\right).
\eeq 
Furthermore, when $n_{\mathrm{cm}}=1$, the coupling strengths for the two terms in $H_1(t)$ reduces, apart from multiplicative constants of order one, to those considered in Sec.~\ref{excesmicro} and shown in Fig.~\ref{figexcess}. Again, a repeat of the numerical calculation with $m_i$ replaced with $m$ shows that these coupling strengths are significant only when $d \lesssim 4.8 \, l_{\mathrm{rel}}$. 

Since all the coupling strengths at possible resonances are either vanishing or only significant at trap distances smaller than a characteristics distance  $d_{\mathrm{mm}} \simeq 4.8 l_{\mathrm{rel}}$, all of the conclusions we derived about the micromotion effect in the previous sections are still true for the more realistic model where the atom is allowed to move. Moreover, one could see the scheme mentioned in Sec.~\ref{bypass} is still possible since the gap between the low lying unperturbed molecular states of the relative Hamiltonian is much larger than the micromotion frequency and hence these states are protected from resonances and the resulting micromotion-induced avoided crossings. 

So the qualitative behavior of the micromotion effect is very much the same for the model considered here, which corresponds to $\omega_0 = \omega_a$, as it is for the simplified model of Sec.~\ref{micromotioncoupling}, which corresponds to $\omega_0\ll \omega_a$. Thus, we expect all the main features of the micromotion effect to occur for ${\omega_0}\lesssim {\omega_a}$. These features include resonances, micromotion-induced avoided crossings, and  the existence of the characteristic distance $d_{\mathrm{mm}}$. In the other regime where ${\omega_0} \gg {\omega_a}$, which means a very large trapping frequency for the ion trap, the ion must be tightly bounded to the center of its trap; so it is possible for the atom trap to come really close to the ion without sensing a very small amplitude of the micromotion. This fact can be seen easily by considering the extreme limit of ${\omega_0}\rightarrow \infty$, for which the trapped ion can be treated as a fixed ion at the center of its trap. In this case, we retrieve the simplified model described in Sec.~\ref{micromotioncoupling} with the atom and the ion interchanging their roles. It is obvious that the micromotion does not enter the Hamiltonian for this model and the exact energy diagrams must look similar to the ones shown in Fig.~\ref{figee}. These observations imply that, for applications such as the quantum controlled phase gate which require the atom trap and the ion trap to be as close as possible, it is better to use a configuration with a large secular frequency ${\omega_0}$.

The precise condition for the trapped ion to be considered as a fixed ion, meaning micromotion would be of no importance, requires the value of the minimum trap distance involved in each specific problem. This minimum trap distance is around $d_c=2R_a^{1/3}l_a^{2/3}$ for the adiabatic quantum phase gate [see Eq.~\eqref{dc}]. At this trap distance, the shortest atom-ion distance is roughly $d_c-l_a -l_i$. For the ion to be treated as a fixed ion, this distance must be much larger than the harmonic oscillator length $l_i$ of the ion, which yields
\beq
2\frac{l_i}{R_a}+\frac{l_a}{R_a}\ll 2\left(\frac{l_a}{R_a}\right)^{2/3}.
\eeq
After interchanging the subscripts $i$ and $a$, one obtains the condition for the trapped atom to be treated as a fixed atom. 

For the numerical calculations in this section we use $\omega_0=\omega_a=2\pi \times 100$ kHz and $\omega=2\pi \times 1.27$ MHz, which are unrealistically small for an ion trap. We repeated our calculation of the coupling strengths for the more reasonable values $\omega_0=\omega_a=2\pi \times 1$ MHz and $\omega=2\pi \times 12.7$ MHz and observed the familiar sharp rising at $d_{\mathrm{mm}}\approx 6.1 \, l_{\mathrm{rel}}$. Hence, the micromotion effect, as we describe it, is still valid for these values of the secular frequency and the micromotion frequency. For typical trapping frequencies encountered in experiments, say $\omega_a=2\pi \times 100$ kHz and $\omega_0=2\pi \times 1$ MHz, we have $l_i\approx 0.5 \, l_a$. Since $l_i$ does not differ greatly from $l_a$, the geometry of this realistic system is closer to that of a symmetric model with $\omega_0=\omega_a$, which is discussed in this section, than it is to a model with $\omega_0 \ll \omega_a$ or  $\omega_0 \gg \omega_a$.

\subsection{Trapped atom-ion systems in 3D}\label{3D}
Again we consider the system illustrated in Fig.~\ref{figtrapatomion}, but now the atom is stored in a 3D harmonic trap and the ion is confined in a linear Paul trap \cite{Berkeland98}. Let us denote the components of the position vectors by $r_{a,k}$ for the atom and $r_{i,k}$ for the ion $(k=x,y,z)$. The driving potential used to confine an ion in a 3D rf trap is
\begin{align}
V_i(t)=\sum_k \frac{1}{2}m_iW_{k}(t)r_{i,k}^2,
\end{align}
where
\beq
W_{k}(t)=\frac{\omega^2}{4}\left[a_{k}+2q_{k}\cos (\omega t)\right],
\eeq 
and the parameters $a_{k}$ and $q_{k}$ obey Eq.~\eqref{paul}. The axial trapping frequency of the Paul trap is $\omega_{i,z}=\omega\sqrt{a}/2$ and the radial secular frequencies are
\begin{align}
\omega_{i,x}=\omega_{i,y}=\omega_0,
\end{align}
where $\omega_0$ is given in Eq.~\eqref{sf}.
By taking into account that the trapping potential for the atom is
\beq
V_a=\sum_k \frac{1}{2}m_a\, \omega_{a,k}^2 \, r_{a,k}^2,
\eeq
one can show that the transformed Hamiltonian for the trapped atom-ion system is the sum
\beq
H(t)=H_a + H_i(t) + V_{\mathrm{int}}
\eeq
of the Hamiltonian of a single trapped atom,
\beq
H_a=\sum_k \left(\frac{P_{a,k}^2}{2m_a}+\frac{1}{2}m_a \, \omega_{a,k}^2 \, r_{a,k}^2\right), 
\eeq
the transformed Hamiltonian of a single trapped ion,
\begin{align}
H_i(t)=&\sum_k \left[\frac{P_{i,k}^2}{2m_i}+\frac{1}{2}m_i \, \omega_{i,k}^2 \left(r_{i,k}-d\right)^2\right] + H_{\mathrm{mm}}(t), \nn \\
H_{\mathrm{mm}}(t)=&- m_i (\gamma\omega_0)^2 (X_i-d)^2 \cos(2 \omega t)\nn \\
&- \gamma\omega_0 \{X_i-d,P_{i,x}\} \sin( \omega t)\nn \\
&- m_i (\gamma\omega_0)^2 (Y_i-d)^2 \cos(2 \omega t)\nn \\
&+ \gamma\omega_0 \{Y_i-d,P_{i,y}\},
\end{align}
and the interaction potential,
\beq
V_{\mathrm{int}}=-\frac{\alpha e^2}{2(\vec{r}_{i}-\vec{r}_{a})^4}.
\eeq

At large trap distances, $V_{\mathrm{int}}$ can be ignored so that the motions of the atom and ion are decoupled and the quasienergies of the system are simply the sum of the atom eigenenergies
\beq
E_a=\sum_k \left(n_{a,k}+\frac{1}{2}\right)\hbar \omega_{a,k}, 
\eeq
and the ion's quasienergies, as mentioned in Eq.~\eqref{qe},
\beq
\epsilon_i = \sum_{k=x,y}\left(n_{i,k}+\frac{1}{2}\right)\hbar \mu_{k} + \left(n_{i,z}+\frac{1}{2}\right)\hbar \omega_{i,z} .
\eeq
This asymptotic energy spectrum can be well described in the harmonic approximation. However, we know that at small trap distances the interaction potential will distort the unperturbed eigenstates and eigenenergies so that they come into resonance with each other. This is when the exact quasienergies differ greatly from the approximate eigenenergies and micromotion-induced avoided crossings appear in the quasienergy diagram. Idziaszek \emph{et al.} investigated the unperturbed Hamiltonian for the cases of quasi-1D elongated traps and spherically symmetric traps in Refs.~\cite{Idziaszek07,Doerk10} and the energy spectrum they obtained is very similar to that shown in Fig.~\ref{figee}. Therefore, we expect micromotion-induced couplings to exhibit the same sudden rise at some characteristic distance $d_{\mathrm{mm}}$ as in Fig.~\ref{figcoupling}. The quasinergy diagram of the 3D system must also look similar to the one obtained in Fig.~\ref{figfullzoneqe}, where the energy spectrum changes completely and micromotion-induced avoided crossings appear at trap distances smaller than $d_{\mathrm{mm}}$. Thus, all the qualitative statements we made for the simplified system considered in Sec.~\ref{micromotioneffect}B should carry over to a realistic 3D system. The values of parameters of interest like the characteristic distance $d_{\mathrm{mm}}$ cannot be computed with high accuracies due to the lack of knowledge about the exact value of the short-range phase and must be determined by experiments.

Because of the asymmetrical nature of the atom and Paul traps, the dynamics of the system in an adiabatic process depends on the direction along which we move the traps. From the form of the driving potential $V_i(t)$ and the micromotion Hamiltonian $H_{\mathrm{mm}}(t)$ we notice that the micromotion occurs only in the radial directions. Along the axial direction the ion behaves like a harmonic oscillator. When the atom and ion are trapped very tightly in the radial direction it is likely that the micromotion effect is weaker if we move the trap along the axial direction. The ideal configuration to minimize the micromotion effect is the one for which the micromotion frequency is as large as possible compared with the radial trapping frequency of the Paul trap, and for both traps the radial frequencies should be much larger than the axial trapping frequencies\cite{Nguyen12}. However, even then the micromotion cannot be ignored at small trap distances because the interaction potential $V_{\mathrm{int}}$ clearly couples the motion along the radial directions to the motion along the axial direction and thus it is possible for the micromotion to be propagated to the motion along the axial direction. This transfer of the micromotion is negligible for relatively large trap distances but it must be significant at small trap distances, especially when the system transforms to a molecular state for which the atom and ion are close to each other.

Our analysis from Sec.~\ref{micromotioneffect}A to Sec.~\ref{micromotioneffect}F implies that applications based on a Floquet adiabatic process, such as the creation of the atom-ion macromolecule and the implementation of the quantum phase gate, are realizable as long as we can carry out the following steps: (i) preparing the trapped ion in its Floquet ground state, (ii) moving the ion trap adiabatically to the atom trap and (iii) make a transition to a low lying molecular state at a trap distance $d_1>d_{\mathrm{mm}}$ before continuing with the adiabatic process until $d \simeq 0$.  Requirement (i) raises the question about whether the micromotion has any effect on the cooling of a trapped ion to its ground state. The laser cooling process of trapped ions when the intrinsic micromotion is taken into account is discussed in Refs.~\cite{Cirac94,Leibfried03}, and the influence of the excess micromotion is studied in Ref.~\cite{Berkeland98}. Even though the micromotion may result in unwanted heating of the ion, cooling can still be achieved by choosing the right value for the laser frequency. Therefore, the intrinsic and the excess micromotion are not fundamental obstacles for the task of cooling a trapped ion to its ground Floquet state.

\section{conclusions}
In this work we show that while the micromotion can be neglected for single trapped ions it must be taken into account when the ion strongly interacts with an external system. In the context of trapped atom-ion interaction, the effect of the  micromotion is most important when there is resonance between two eigenstates of the approximate Hamiltonian. This happens when the energy separation is close to the micromotion frequency and the coupling that arises from the micromotion between the two states is significant. The micromotion leads to not only a shift in energy but also numerous new avoided crossings in the energy diagram. The magnitude of these effects can be inferred by (i) a calculation of micromotion-induced coupling strengths and (ii) an exact computation of the quasienergies of the system. Although the second method gives much more detail, the first method is much faster yet sufficiently accurate to predict the main features of the quasienergy diagram. Our results show that the approximation that the ion trap is a harmonic trap breaks down at trap distances smaller than a characteristic distance $d_{\mathrm{mm}}$. The quasienergy diagram for this region is far more complicated than the approximate eigenenergy diagram and thus the behavior of the system in an adiabatic process may change completely when the micromotion is included. The complication due to the presence of the micromotion may seriously reduce the fidelity of the proposed atom-ion quantum gate. However, we suggest a scheme to overcome this difficulty by utilizing a transition from the initial ground Floquet state to a low-lying molecular state before the onset of the micromotion effect. These transitions can also be used to create an atom-ion macromolecule which may indeed be of interest in its own right.

\begin{acknowledgements}
We are grateful to T. Calarco and Z. Idziaszek for valuable suggestions and helpful comments. We also thank P. Zoller, P. Rabl, and S. Habraken for useful discussions and their kind hospitality at the Institute for Quantum Optics and Quantum Information (IQOQI). This work is supported by the NUS Graduate School for Integrative Sciences and Engineering (NGS) and the Centre for Quantum Technologies (CQT). CQT is a Research Centre of Excellence funded by Ministry of Education and National Research Foundation of Singapore.   
\end{acknowledgements}

\appendix
\section{Adiabatic quantum phase gate in the Floquet picture}\label{floquetadiabatic}
Before we discuss the adiabatic quantum phase gate in the Floquet picture, we need to understand how a quantum system evolves in a Floquet adiabatic process \cite{Drese99}. We start with the Schr\"odinger equation
\beq
i\hbar \frac{\partial}{\partial t} \Psi(t) = H\big(\lambda(t),t\big)\Psi(t),
\eeq
where the parameter $\lambda$, which is the trap distance in the specific case of the trapped atom-ion system, is varied slowly in time. At each moment we have the instantaneous Floquet states which satisfy the Floquet eigenvalue equation
\beq
 H_F(\lambda ,t)u_{n,k}(\lambda , t)= \epsilon_{n,k}(\lambda) u_{n,k}(\lambda, t).
\eeq
To study the adiabatic evolution, one needs to employ a trick called the $(t,t')$ method \cite{Peskin93} in which we change the implicit time parameter $t$ in $\lambda(t)$ to a new time variable $t'$ and introduce a higher-dimensional wave function $\tilde{\Psi}(t,t')$ with the requirement that this wave function satisfies the ``extended" Schr\"odinger equation
\beq\label{fsc}
i\hbar \frac{\partial}{\partial t'}\tilde{\Psi}(t,t')=H_F\big(\lambda(t'),t\big)\tilde{\Psi}(t,t').
\eeq
When the initial condition of  $\Psi(t=0)$ is transformed to that of $\tilde{\Psi}(t,t'=0)$ in an appropriate way \cite{Drese99}, we have 
\beq
\Psi(t)=\tilde{\Psi}(t,t'=t). 
\eeq
An advantage of using the extended wave function $\tilde{\Psi}(t,t')$ is the similarity of Eq.~\eqref{fsc} with the Schr\"odinger equation encountered in the normal adiabatic process for which the Hamiltonian only depends on time implicitly through the parameter $\lambda$, and thus well-known perturbation techniques can be used. In this picture the variable $t$ is treated as a fourth dimension of space and hence can be dropped from now on. 

We expresses the wave function $\tilde{\Psi}(t')$ in terms of the instantaneous Floquet states,
\begin{align}
\tilde{\Psi}(t')=&\sum_{n,k} c_{n,k}(t') u_{n,k}\big(\lambda(t')\big)\nn \\
&\times \exp \left[-\frac{i}{\hbar} \int_{0}^{t'}d\tau \, \epsilon_{n,k}\big(\lambda(\tau)\big)\right],
\end{align}
and insert it into Eq.~\eqref{fsc} to obtain the differential equations for the coefficients $c_{n,k}(t')$ which in turn can be solved approximately by perturbation methods. If the system starts in a single Floquet state, say $u_{0,0}$, then first-order perturbation theory gives us
\begin{align}
c_{n,k}(t')\!=&-\int^{t'}_{0} d\tau \frac{\partial \lambda(\tau)}{\partial \tau} \nn \\
&\times \left\langle \left\langle u_{n,k}\big(\lambda(\tau) \big)\left|\frac{\partial}{\partial \lambda}\right|u_{0,0}\big(\lambda(\tau)\big)\right\rangle \right\rangle \nn \\
&\times\! \exp \left\{\frac{i}{\hbar} \int_{0}^{\tau}\!\! d\tau'\! \left[\epsilon_{n,k}\big(\lambda(\tau')\big)-\epsilon_{0,0}\big(\lambda(\tau')\big)\right]\right\}.
\end{align}

Recall that all the Floquet states $u_{n,k}$ correspond to a unique physical state $u_n$, to find the transition amplitude to the physical state $u_n$ at the real time $t$ we need to sum over all values of $k$ and then replace $t'$ with $t$
\beq
c_{n}(t)=\sum_{k=-\infty}^{\infty} c_{n,k}(t'=t).
\eeq
For the system to almost stay in a single Floquet state at any moment in time, we need $\left|c_n(t)\right|\ll 1$ for all $n$, which is the general requirement for a Floquet adiabatic process. 

Let us now consider the Floquet adiabatic process used to implement the quantum phase gate as considered in Sec.~\ref{micromotioncoupling}. The adiabatic theorem states that if the initial motional state is a Floquet state $u_n(d(t_i),t_i)$ of the system, the final motional state at the end of the process is 
\beq
\phi(t_f)=u_n\big(d(t_f), t_f\big)\exp\left[-\frac{i}{\hbar}\int^{t_f}_{t_i} d\tau \, \epsilon_n\big(d(\tau)\big)\right],
\eeq
where it is understood that the quasienergy of the ground state is set to zero at all trap distances. Since the instantaneous quasienergy $\epsilon_n(d)$ depends on the interaction potential which in turn depends on the spin of the atom and the ion, we know that the phase accumulated in an adiabatic process must be spin dependent \cite{Doerk10}. 

In the spin subspace, one can create, by single qubit gates, a separable state
\beq
\ket{\chi(t_i)}=\sum_{k=1}^4  \ket{k} c_k,
\eeq
with $k = 00, 01, 10, 11,$ and $c_1 c_4=c_2 c_3$. Again ``spin" means the hyperfine levels of the atom and the ion which are chosen as the qubit states. The Floquet adiabatic process turns the initial state,
\beq \label{ipsi}
\ket{\Psi(t_i)}=u_0(d(t_i),t_i)\sum_{k=1}^4  \ket{k} c_k,
\eeq
to
\beq
\ket{\Psi(t_f)}=u_0(d(t_f),t_f)\sum_{k=1}^4 \ket{k}c_k e^{i\theta_k}.
\eeq
Each of the spin terms $\ket{k}$ accumulates a different phase $\theta_k$ since the phase in an adiabatic process is spin dependent. The final spin state is $\ket{\chi(t_f)}=\sum_{k=1}^4 \ket{k} c_k e^{i\theta_k}$, and the adiabatic process is equivalent to the application of a quantum phase gate $U=\mathrm{diag} \{e^{i\theta_k}\}$ in the spin subspace.

In a realistic situation there must be a small fraction of the motional state being transferred to excited Floquet levels at the end of the adiabatic process. We consider the simplified situation when only one dominant excited Floquet state $u_e$ is occupied with a small probability $p_e$. If we first ignore the spin degree of freedom and write the amplitude of the ground state and the excited state in their polar forms, the final motional state, apart from a global phase, is 
\beq
\bar{\phi}(t_f)=\sqrt{p_0} u_0\big(d(t_f),t_f\big)e^{i\theta}+\sqrt{p_e} u_e\big(d(t_f),t_f\big)e^{i{\theta}'},
\eeq
with $p_0+p_e=1$ and $p_e\ll 1$. 

For the initial state given in Eq.~\eqref{ipsi}, because of the spin dependence of the phases $\theta$ and $\theta'$, the final total wave function is more complicated,
\beq
\ket{\bar{\Psi}(t_f)}=\sum_{k=1}^4 \ket{\varphi_k}\ket{k} c_k e^{i\theta_k},
\eeq
where
\beq
\ket{\varphi_k}=\sqrt{p_0} u_0\big(d(t_f),t_f\big)+\sqrt{p_e} u_e\big(d(t_f),t_f\big)e^{i\alpha_k},
\eeq
with $\alpha_k={\theta}'_k-\theta_k$.
Note that in the above we have ignored a very small spin-dependence in the amplitude $\sqrt{p_0}$ and $\sqrt{p_e}$. 

A comparison of the realistic final wave function $\ket{\bar{\Psi}(t_f)}$ and the ideal wave function $\ket{\Psi(t_f)}$ shows that when branching occurs the motional state is entangled to the spin state, and the adiabatic process no longer results in a pure state in the spin subspace. The density matrix for the mixed state of the spin subspace can be obtained by taking the partial trace of the total density matrix $\ket{\bar{\Psi}(t_f)}\bra{\bar{\Psi}(t_f)}$; the result is
\beq
\rho(t_f)=\sum_{k,j}  \ket{k}c_k\lambda_{kj}e^{i(\theta_k-\theta_j)}c^*_j \bra{j},
\eeq
where
\beq
\lambda_{kj}=\braket{\varphi_j | \varphi_k}=1-p_e\left[1-e^{i(\alpha_k-\alpha_j)}\right].
\eeq
The gate fidelity is defined as \cite{Nielsen00}
\beq
F= \min \left\{\sqrt{\bra{\chi(t_f)}\rho(t_f)\ket{\chi(t_f)}}\right\},
\eeq
where the minimization is done over all possible values of $c_k$ satisfying the constraint $\sum_k|c_k|^2=1$ . 

Now we introduce the $4 \times 4$ symmetric matrix $M$ whose elements are
\beq
M_{kj}=1-\cos(\alpha_k-\alpha_j),
\eeq
and the column vector $V$ with the components $V_j=|c_j|^2$. It follows that
\beq
\bra{\chi(t_f)}\rho(t_f)\ket{\chi(t_f)}=1-p_eV^T M V,
\eeq
and the maximum value of $V^T M V$, subjected to the constraint $\sum_k|c_k|^2=1$, is
\beq
\max \left\{V^T M V\right\} = \left(A^TM^{-1}A\right)^{-1},
\eeq
with the constant vector $A^T=(1,1,1,1)$. Therefore, the gate fidelity is
\beq
F=\sqrt{1-p_e\left(A^TM^{-1}A\right)^{-1}}\approx 1-\frac{p_e}{2}\left(A^TM^{-1}A\right)^{-1}.
\eeq
There is no simple analytic form for $A^TM^{-1}A$; however, using the fact that $M_{kj}=0$ for $k=j$ and $M_{kj}\leq 2$ for $k\neq j$, one can show that $V^T M V\leq \frac{3}{2}$ and derive a lower bound for the gate fidelity
\beq
F\geq\sqrt{1-\frac{3}{2}p_e}\approx 1-\frac{3}{4}p_e.
\eeq
In short, we see that
$
1-F\propto p_e
$.
Thus, the gate fidelity decreases linearly with the occupancy of the excited Floquet state.

\section{Floquet state of a single trapped ion}\label{fssti}

Here we consider a single ion confined in a rf trap with both the intrinsic micromotion and the excess micromotion. The Floquet state of an ion with only the intrinsic micromotion is given by Glauber in Ref.~\cite{Glauber07}. Here we apply his elegant proof to derive the Floquet state of a trapped ion in the presence of the excess micromotion. The potential for the motion along the $x$ axis is
\begin{align}
V_x(t)=&\frac{1}{8}m_i\omega^2\left[a+2q \cos \omega t\right]X^2 \nn \\
&- e\left[E_{\mathrm{dc}}+E_{\mathrm{ac}}\sin (\omega t)\right]X,
\end{align}
and the equation of motion is an inhomogeneous differential equation:
\beq\label{iheq}
\ddot{X}(t)+\left[a+2q\cos (\omega t)\right] \frac{\omega^2}{4}X(t)=F_0 + F_1 \sin (\omega t),
\eeq
where $F_0 = eE_{\mathrm{dc}}/m_i$, and $F_1=eE_{\mathrm{ac}}/{m_i}$.

The homogeneous equation is the Mathieu equation which possesses a special Floquet solution $f(t)$ satisfying the initial condition
\beq
f(0)=1, \ \dot{f}(0)=i\nu.
\eeq
This special solution has the expression
\beq\label{fsol}
f(t)=e^{i\mu t} \varphi(t),
\eeq
where $\mu$ is called the Floquet exponent which depends on $a,q,\omega$ and $\varphi(t)$ is a periodic function whose Fourier series is
\beq\label{phi}
\varphi(t)=\sum_n C_{n}e^{i n\omega t}.
\eeq
The coefficients $C_n$ and the Floquet critical exponent $\mu$ can be found by simple numerical procedures \cite{Stegun65}. The value of $\nu$ can be then computed from the relation
\beq
\nu = \mu +\omega \sum_{n} n C_n.
\eeq
When $|a|, |q| \ll 1$, the lowest-order approximation in $a$ and $q$ gives 
$
\mu \approx \omega_0,
$
with $\omega_0$ defined in Eq.~\eqref{sf}, and \cite{Leibfried03}
\beq
\varphi(t)\approx\frac{1+(q/2)\cos (\omega t)}{1+q/2}. 
\eeq

If $f(t)$ is the solution, $f(t)^*$ must also be a solution, and a general homogeneous solution has the form
\beq
x_h=Af(t)+Bf(t)^*.
\eeq
When $|a|, |q| \ll 1$ we have $\mu \ll \omega$ so the Floquet exponent cannot be an integer multiple of $\omega$. This means that $f(t)$ and $f(t)^*$ are not periodic functions and hence no homogeneous solution is periodic. 

We now show that the inhomogeneous equation \eqref{iheq} has one and only one periodic solution when $|a|, |q| \ll 1$. It is easy to see that there is not more than one periodic solution: Assume $x_p(t)$ is a particular periodic solution; any other solution $x'_p(t)$ is the combination of $x_p(t)$ and a homogeneous solution, that is,
\beq
x'_p(t)=x_p(t)+Af(t)+Bf(t)^*.
\eeq 
Since $f(t)$ and $f(t)^*$ are not periodic, it is impossible for $x_p'(t)$ to be periodic unless $A=B=0$, which results in $x_p'(t)=x_p(t)$. Thus, we conclude that the inhomogeneous equation has no more than one periodic solution. 

We prove the existence of the unique periodic solution by explicitly constructing it by the variation-of-parameters method \cite{Riley06}. We look for a periodic solution of the form
\beq
x_p(t)=k(t)f(t)+k(t)^*f(t)^*
\eeq
subject to the condition
\beq
\dot{k}(t) f(t)+\dot{k}(t)^*f(t)^* = 0.
\eeq
By making use of the time-independent Wronskian,
\beq
\dot{f}(t) f(t)^*-\dot{f}(t)^*f(t)=2i\nu,
\eeq
we obtain 
$
x_p(t)=\Re\{\Phi(t) \varphi(t)^*\}
$,
where $\varphi(t)$ is the function given in Eq.~\eqref{phi} and 
\beq
\Phi(t)=\sum_{n=-\infty}^{\infty}D_n e^{i n\omega t},
\eeq
with
\beq
D_n=\frac{2C_nF_0 + iF_1(C_{n+1}-C_{n-1})}{2\nu(\mu+n\omega)}.
\eeq
In the lowest-order approximation we have
\beq
x_p(t)\approx {\delta d}\left[1+\frac{q}{2}\cos (\omega t)\right]-\frac{\omega_0}{\omega}l_{\mathrm{ac}}\sin (\omega t),
\eeq
where $\delta d$ and $l_{\mathrm{ac}}$ are defined in Sec.~\ref{excesmicro} and $\omega_0$ is defined in Eq.~\eqref{sf}. It is obvious that $x_p(t+T)=x_p(t)$ and so we have found the unique periodic solution of Eq.~\eqref{iheq}.

Glauber's elegant approach can be used to find the quasienergies and Floquet states of the trapped ion. First we introduce a new position operator that depends explicitly on time, 
\beq\label{X1}
X_1(t)=X(t)-x_p(t),
\eeq
so that $X_1(t)$ satisfies the homogeneous differential equation. The Wronskian 
\beq
f(t)\dot{X}_1(t)-\dot{f}(t)X_1(t)
\eeq
is constant in time and thus the operator
\beq\label{A}
A=i\sqrt{\frac{m_i}{2\hbar \nu}} \left[f(t)\dot{X}_1(t)-\dot{f}(t)X_1(t)\right]
\eeq
is also a constant of motion, and it possesses constant eigenkets. The operator $A$ and its adjoint are the analogs of the ladder operators for the quantum harmonic oscillator, and one can show that $[A,A^{\dag}]=1$. Let us define the states
\begin{align}
A\ket{0}&=0, \nn \\
\ket{n}&=\frac{\left(A^{\dag}\right)^n}{\sqrt{n!}}\ket{0},
\end{align}
and demonstrate that they are indeed the Floquet states of the trapped ion. 

Although the eigenket $\ket{0}$ is time independent, its representing wave function $\braket{{x,t|}0}$ is not. We have $\bra{x,t}A\ket{0}~=~0$, which yields
\begin{align}
\left\{\frac{\hbar}{im_i}f(t)\frac{\partial}{\partial x}-\dot{f}(t)\left[x-x_p(t)\right]-\dot{x}_p(t)f(t)\right\}\braket{{x,t|}0}=0,
\end{align}
The solution in its normalized form is
\begin{align}
\braket{{x,t|}0}=&\left(\frac{m_i \nu}{\pi \hbar}\right)^{1/4}\frac{e^{is(x,t)}}{\left[f(t)\right]^{1/2}}\nn \\
&\times \exp \left\{\frac{i m_i}{2\hbar}\frac{\dot{f}(t)}{f(t)}\left[x-x_p(t)\right]^2\right\},
\end{align}
with
$
s(x,t)=m_i\dot{x}_p(t)x/\hbar. 
$
By applying the operator $A^{\dag}$ repeatedly to the state $\ket{0}$ we obtain
\begin{align}
\braket{{x,t|}n}=&\frac{1}{\sqrt{2^n n!}}\left(\frac{m_i \nu}{\pi \hbar}\right)^{1/4}\frac{e^{is(x,t)}}{[f(t)]^{1/2}}\left[\frac{f(t)^*}{f(t)}\right]^{n/2} \nn \\
&\times H_n\left(\left[\frac{m_i \nu}{\hbar |f(t)|^2}\right]^{1/2}[x-x_p(t)]\right) \nn \\
&\times \exp \left\{\frac{im_i}{2\hbar}\frac{\dot{f}(t)}{f(t)}\left[x-x_p(t)\right]^2\right\},
\end{align}
where $H_n$ is the Hermite polynomial of order $n$. Upon inserting $f(t)$ from Eq~\eqref{fsol} to the above expression, we get
\beq
\braket{x,t|n}=\exp \left[-i\left(n+\frac{1}{2}\right)\mu t \right] u_n(t),
\eeq
with the periodic functions
\begin{align}
u_n(t)=&\frac{1}{\sqrt{2^n n!}}\left(\frac{m_i \nu}{\pi \hbar}\right)^{1/4}\frac{e^{is(x,t)}}{[\varphi(t)]^{1/2}}\left[\frac{{\varphi}(t)^*}{\varphi(t)}\right]^{n/2}\nn \\
&\times H_n\left(\left[\frac{m_i \nu}{\hbar |f(t)|^2}\right]^{1/2}[x-x_p(t)]\right) \nn \\
&\times \exp \left\{-\frac{m_i\mu}{2\hbar}\left[1-\frac{i\dot{\varphi}(t)}{\mu\varphi(t)}\right]\left[x-x_p(t)\right]^2\right\}.
\end{align}
 
The structures of the wave functions $\braket{{x,t|}n}$ suggest they are the Floquet states with quasienergies 
\beq
\epsilon_n=\left(n+\frac{1}{2}\right)\hbar \mu.
\eeq
Since $\mu$ is independent of $E_{\mathrm{dc}}$ and $E_{\mathrm{ac}}$, the quasienergies of a single trapped ion are indeed not affected by the excess micromotion. 

In situations where the excess micromotion vanishes, we have $x_p(t)=0$ and the wave functions of the Floquet states $\ket{n}$ reduce to the forms obtained by Glauber in Ref.~\cite{Glauber07} as expected. A large amount of the excess micromotion, which results in a large amplitude of $x_p(t)$, will lead to strong oscillations in the wave functions $u_n(t)$. 

Let us now work out the kinetic energy of the trapped ion when it is prepared in the state $\ket{n}$. Using Eqs.~\eqref{X1} and \eqref{A}, we have
\beq
P(t)=m_i\dot{X}(t)=\sqrt{\frac{\hbar m_i}{2\nu}}\left(\dot{f}(t)^*A+\dot{f}(t)A^{\dag}\right)+m_i\dot{x}_p(t),
\eeq
and the kinetic energy associated with the state $\ket{n}$ is 
\beq
\frac{\bra{n}P(t)^2\ket{n}}{2m_i}=\frac{1}{2}\left(n+\frac{1}{2}\right)\hbar\frac{|\dot{f}(t)|^2}{\nu}+\frac{m_i[\dot{x}_p(t)]^2}{2}.
\eeq
The mean kinetic energy is defined as the time-averaged value $\overline{\braket{P(t)^2}}/2m_i$ taken over one period $2\pi/\omega$ of the micromotion. After inserting $f(t)$ from Eqs.~\eqref{fsol} and \eqref{phi} to the above equation and taking the time average we arrive at
\begin{align}
\frac{\overline{\bra{n}P(t)^2\ket{n}}}{2m_i}=&\frac{1}{2}\left(n+\frac{1}{2}\right)\frac{\hbar}{\nu}\sum_{n=-\infty}^{\infty}C_n^2(\mu+n\omega)^2\nn \\
&+\frac{m_i\overline{\dot{x}_p(t)^2}}{2},
\end{align}
and when $|a|\ll q^2 \ll 1$ we have 
\beq
\frac{\overline{\bra{n}P(t)^2\ket{n}}}{2m_i}\approx \Big(n+\frac{1}{2}\Big)\hbar\omega_0+\frac{1}{2}m_i\omega_0^2\left[\left(\delta d\right)^2+l_{\mathrm{ac}}^2/2\right].
\eeq
For the ground state we have
\beq 
\frac{\overline{\braket{P(t)^2}}}{2m_i}\approx \frac{\hbar\omega_0}{2}+\frac{1}{2}m_i\omega_0^2\left[\left(\delta d\right)^2+l_{\mathrm{ac}}^2/2\right].
\eeq
The cooling of a trapped ion to its ground Floquet state and the final value of its mean kinetic energy are studied in Ref.~\cite{Cirac94} (without the excess micromotion). We see that the excess micromotion leads to an increase in the mean kinetic energy of the ground state, but this does not mean a worse cooling limit if we are only interested in the population of the ground state.

\end{document}